\documentclass[twocolumns]{aa} %

\usepackage{graphicx}
\usepackage{epstopdf}
\usepackage{deluxetable}
\usepackage{times,epsfig} 
\usepackage{mathrsfs}  
\usepackage{natbib}
\usepackage{amssymb}
\bibpunct{(}{)}{;}{a}{}{,} 
\usepackage[english]{babel}
\usepackage{longtable}
\usepackage[english]{babel}
\usepackage{url}
\usepackage{hyperref}
\usepackage{color}
\usepackage[normalem]{ulem}

\begin{document}

\title{Late-time observations of the extraordinary Type II supernova iPTF14hls}

\author
{
J. Sollerman\inst{1},
F. Taddia\inst{1},
I. Arcavi\inst{2,3,4,5}, 
C. Fremling\inst{1,6},  
C. Fransson\inst{1},    
J. Burke\inst{2,3},      
S.~B. Cenko\inst{7},   
O. Andersen\inst{1},    
I. Andreoni\inst{6,8},    
C. Barbarino\inst{1},   
N. Blagorodova\inst{6}, 
T. G. Brink\inst{9},       
A.~V. Filippenko\inst{9,10}, 
A. Gal-Yam\inst{11},       
D. Hiramatsu\inst{2,3},     
G. Hosseinzadeh\inst{2,3}, 
D. A. Howell\inst{2,3},   
T. de Jaeger\inst{9},   
R. Lunnan\inst{1},    
C. McCully\inst{2,3},  
D. A. Perley\inst{12},      
L. Tartaglia\inst{1},   
G. Terreran\inst{13},  
S. Valenti\inst{14},    
X. Wang\inst{15} 
}
      
\institute{The Oskar Klein Centre, Department of Astronomy, Stockholm University, AlbaNova, 10691 Stockholm, Sweden
\and Department of Physics, University of California, Santa Barbara, CA 93106-9530, USA
\and Las Cumbres Observatory, 6740 Cortona Dr Ste 102, Goleta, CA 93117-5575, USA
\and The Raymond and Beverly Sackler School of Physics and Astronomy, Tel Aviv University, Tel Aviv 69978, Israel
\and Einstein Fellow
\and Division of Physics, Math and Astronomy, California Institute of Technology, 1200 East California Boulevard, Pasadena, CA 91125, USA ; Cahill Center for Astrophysics, California Institute of Technology, Pasadena, CA 91125, USA
\and Astrophysics Science Division, NASA Goddard Space Flight Center, Mail Code 661, Greenbelt, Maryland 20771, USA
\and Centre for Astrophysics and Supercomputing, Swinburne University of Technology, PO Box 218, Hawthorn, 3122, VIC, Australia
\and Department of Astronomy, University of California, Berkeley, CA, 94720-3411, USA
\and Miller Senior Fellow, Miller Institute for Basic Research in Science, University of California, Berkeley, CA  94720, USA
\and Benoziyo Center for Astrophysics, Weizmann Institute of Science, Rehovot 76100, Israel
\and Astrophysics Research Institute, Liverpool John Moores University, IC2, Liverpool Science Park, 146 Browlow Hill, Liverpool L3 5RF, UK
\and Center for Interdisciplinary Exploration and Research in Astrophysics (CIERA) and Department of Physics and Astronomy, Northwestern University, Evanston, IL 60208, USA
\and  Department of Physics, University of California, 1 Shields Avenue, Davis, CA 95616-5270, USA 0000-0001-8818-0795, USA
\and Physics Department and Tsinghua Center for Astrophysics (THCA), Tsinghua University, Beijing 100084, China
}

\date{Received; accepted}

\abstract
{}
{We study iPTF14hls, a luminous and extraordinary long-lived Type II
  supernova, which lately has attracted much attention and disparate interpretation.}
{We have presented new optical photometry that extends the light
  curves up to more than three~years past discovery. We also obtained optical
  spectroscopy over this period, and furthermore present additional space-based observations using \textit{Swift} and \textit{HST}.}
{After an almost constant luminosity for hundreds of days, the later light curve of iPTF14hls finally fades 
and then displays a dramatic drop after about 1000~d, 
but the supernova is still visible at the latest epochs
  presented. The spectra have finally turned nebular, and our very last
  optical spectrum likely displays signatures from the deep and dense
  interior of the explosion. A high-resolution \textit{HST} image highlights
  the complex environment of the explosion in this low-luminosity galaxy.}
{We provide a large number of additional late-time observations of
  iPTF14hls, which are (and will continue to be) used to assess the many different interpretations for this intriguing object.
{In particular, the very late (+1000 d) steep decline of the optical light curve is difficult to reconcile with the proposed central engine models. The lack of very strong X-ray emission, and the emergence of intermediate-width emission lines including [S II] that we propose originate from dense, processed material in the core of the supernova ejecta, are also key observational tests for both existing and future models.}
}

\authorrunning{Sollerman et al.}
\titlerunning{Late observations of supernova iPTF14hls}


\maketitle
\section{Introduction}
\label{sec:intro}

The extraordinary supernova (SN) iPTF14hls is a Type II supernova (SN~II), and was
discovered and reported by \citet[][hereafter A17]{Arcavi17} as having a 
luminous and long-lived (600+~d) light curve (LC) showing at least five episodes of rebrightening.
The spectra were similar to those of other hydrogen-rich supernovae (SNe),
but evolved at a much slower pace. A17 describe a scenario where this
could be the explosion of a very massive star that ejected a huge
amount of mass prior to explosion. They connect such  eruptions
with the pulsational pair-instability mechanism, but remained cautious
regarding the final interpretation of this highly unusual transient.

Following the report of A17, a number of researchers quickly offered their interpretations of this object. \cite{Chugai18} add to the
 interpretation of A17 and generally agree on the massive ejection
 scenario, while \cite{AS17}
 use a late-time spectrum with a narrow emission line to
 argue for interaction with the circumstellar medium (CSM) as the
 source for the multiple rebrightenings in the LC.
 \cite{Dessart18} instead suggested a magnetar as the powering
 mechanism, whereas \cite{Soker18} prefer a common-envelope jet.
 \cite{WangWangWang18} propose a fall-back accretion model for
 iPTF14hls and \cite{Woosley18} discuss pros and cons of several of
 the above-mentioned models.
 This selection of interpretations for iPTF14hls this includes a large variety of models. 
 More data could help to differentiate between these scenarios.

In this paper, we present a comprehensive set of additional late-time
 observations for iPTF14hls. This extends the optical LC
 monitoring up to more than 3~yr (1236~d in the rest frame) past discovery, 
 which doubles the duration of the LC previously discussed. Additional 
late-time spectroscopy from 
a number of larger telescopes
reveal how iPTF14hls finally enters the nebular phase. 
In addition to the ground-based data, we have
also triggered the \textit{Neil Gehrels Swift Observatory} \citep[\textit{Swift,}][]{gehrels2004} and the \textit{Hubble Space Telescope} (\rm{HST}), and describe these data here.

The paper is structured as follows. 
In Sect.~\ref{sec:data}, the new ground-based optical SN imaging observations
and the corresponding data reductions are presented,
whereas in 
Sect.~\ref{sec:space} we describe the \textit{Swift} and \textit{HST} observations.
The LCs in the different bands are presented and
analysed in conjunction with the data already presented by A17 in Sect.~\ref{sec:lc}, and 
Sect.~\ref{sec:spectra} presents our new series of SN spectra. 
A discussion of the results, in relation to many of the
suggested models in the literature, is given in Sect.~\ref{sec:discussion}. 
{We summarise the discussion in Sect.~\ref{sec:summary}.}

We follow A17 and adopt a redshift of $z=0.0344$, corresponding to a
luminosity distance of 156 Mpc. We correct all photometry for Milky
Way extinction ($E(B-V)=0.014$~mag), but make no correction for host-galaxy
extinction, since we do not detect any narrow \ion{Na}{i}~D at the host-galaxy 
rest wavelength in the spectra. 

\section{Ground-based imaging observations and data reduction}
\label{sec:data}

The optical LCs of iPTF14hls are presentend in A17 from discovery to 
600~d past discovery. Their photometry came from the intermediate Palomar Transient Factory (iPTF) and the 
Las Cumbres Observatory (LCO), and we here report on the continued monitoring of iPTF14hls
with these telescopes, as well as with 
additional telescopes at later epochs. 
In total we present data from discovery until almost 1300~d past discovery.

The iPTF first detected
iPTF14hls on 2014 September 22.53 UT (universal time is used throughout this paper),
using the 48-inch Samuel Oschin telescope
(P48) at Palomar Observatory. This is equivalent to 
JD 2,456,923.03, which we used as the reference date throughout the paper.
Follow-up observations of iPTF14hls on Palomar mountain were obtained with the 
Palomar 60-inch telescope (P60; \citealp{cenko06}). 
Data using P48 and P60 for the first 600~d were reported by A17 and are also included in our LC figures. Past 600~d the object was too faint for P48, but we continued monitoring iPTF14hls with P60 
in $gri$. The SN was detected for the last time with P60 at 875~d. 
While the earlier P60 data were obtained with the GRBcam, 
our new P60 data were taken with the SEDM rainbow camera \citep{SEDM}. 
The P60 data were reduced with \texttt{FPipe} \citep{fremling16}, making use of Sloan Digital Sky Survey (SDSS; \citealp{ahn14}) templates for the host-galaxy subtraction. SDSS stars in the field of iPTF14hls were used as standards to calibrate the P60 photometry. 

Much of the photometry in A17 came from the LCO 
monitoring campaign, which we also continued past 600~d. iPTF14hls was observed with both the 1m and 2m telescopes available at LCO, equipped with $BgVri$ filters. 
The SN was detected for the last time with the LCO-2~m telescope at 
982~d.  
The LCO photometry was reduced using the LCO pipeline \citep{valenti} 
which includes point-spread-function (PSF) photometry and modelling of the host 
background.

\begin{figure}
\includegraphics[width=9cm]{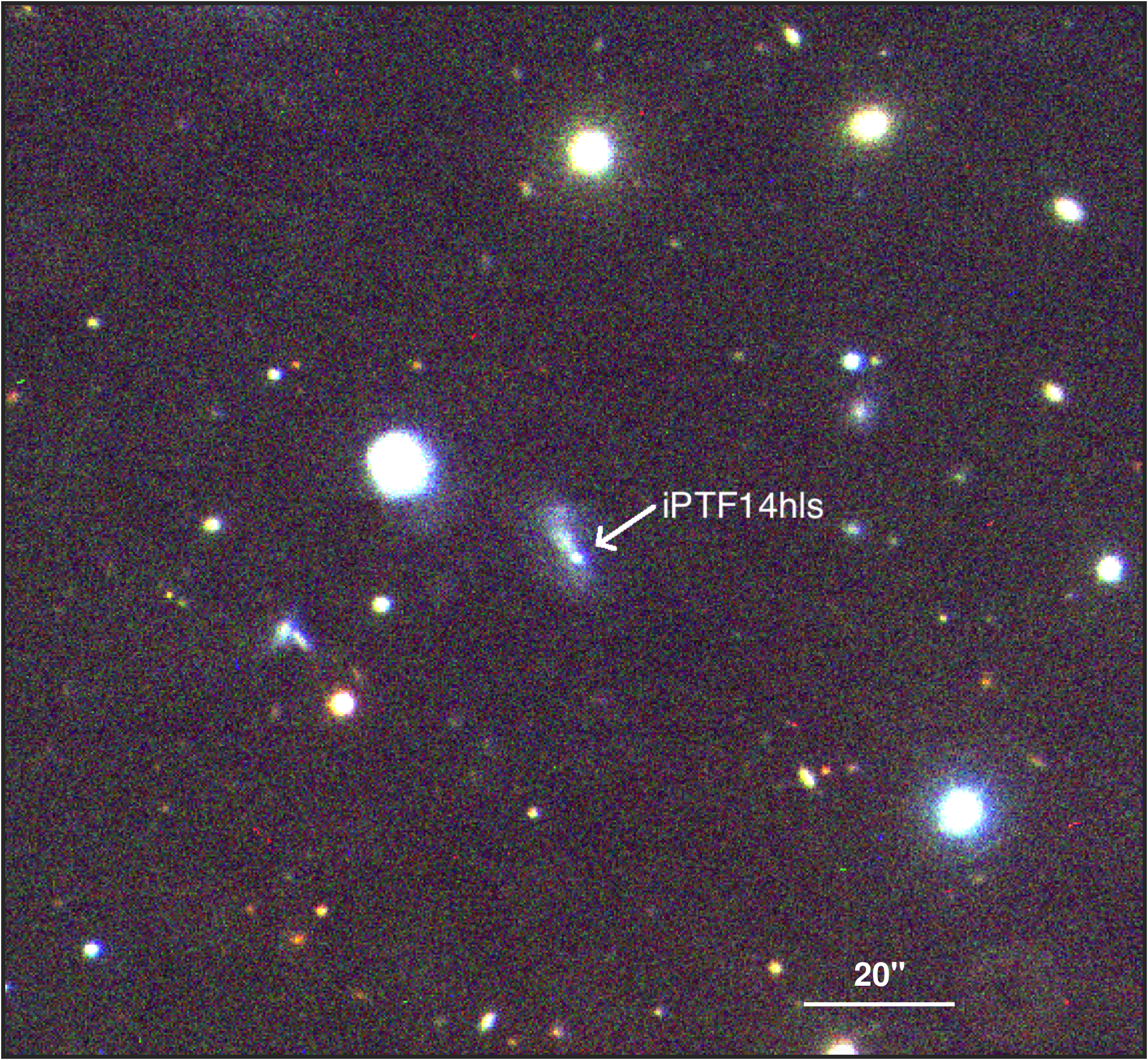}
\caption{\label{NOT} 
Late-time $gri$ composite of SN iPTF14hls taken using the NOT on 17 October 2017, which is 1083 rest-frame days past discovery.
Marked by an arrow is the clearly visible SN in the host galaxy.  North is up and east to the left. The field of view is $\sim 2\farcm53 \times 2\farcm33$.}
\end{figure}

Finally, as the SN continued to fade, and the smaller telescopes were no longer able to detect a significant signal, we obtained imaging with the 2.56~m Nordic Optical Telescope (NOT) using ALFOSC (Andalucia Faint Object Spectrograph and Camera) as well as with the 3.5~m Telescopio Nazionale Galileo (TNG) equipped with DOLORES (Device Optimised for the LOw RESolution), situated next to each other on La Palma. The photometry from NOT and TNG was also reduced with \texttt{FPipe}, using templates and photometric reference stars from SDSS.  
In Fig.~\ref{NOT}, we present a late-time image of the SN and its host galaxy obtained at the NOT. 
A log of all the late-time photometric observations is given in Table~\ref{tab:phot}, and the LCs are presented in Fig.~\ref{lc}.

\section{Space-based observations}
\label{sec:space}

We applied for time to image SN iPTF14hls 
with \textit{HST} in order to investigate the
low-luminosity host in detail and the immediate environment of
iPTF14hls in particular. We were awarded one orbit in Cycle 25\footnote{HST Proposal GO-15222; PI Arcavi.}. 

The main motivation to trigger \textit{Swift} at late epochs was to test the
different scenarios put forward suggesting the existence of a massive
CSM shell around the SN. CSM interaction can be efficiently
probed by radio or X-ray observations.

\subsection{HST}
\label{sec:hst}
\textit{HST} imaging of iPTF14hls was performed in two filters,
F475W and F625W, using the Wide Field Camera 3 (WFC3) and UVIS2. 
 The images were obtained on 2017 December 20, using a three-point dither pattern and a total exposure time 
of 1185~s in each of the two filters. The images were immediately made public and are available in the STSCI archive.
We retrieved the calibrated, geometrically-corrected, dither-combined image files from the archive and the combined image is displayed in Fig.~\ref{HST}.

{We obtained astrometric and photometric measurements from the HST images using a pipeline written by C. McCully which utilises {\tt SEP} \citep{sep}, {\tt Astropy} \citep{astropy}, and {\tt Dolphot}
\cite[see][]{dolphot}.}
The best-fit position (J2000.0) is\\
{RA = 09$^{h}$20$^{m}$34.291$^{s}$, 
Dec = +50$^{d}$41'46.768".} 
                                                     
If future high-resolution images of the host galaxy are obtained, this position can be useful for studies of the environment of the SN explosion. 
A17 characterised the host as a moderately metal-poor galaxy having a mass comparable to that of the Small Magellanic Cloud (SMC). The physical extent of the host, as measured in the {\it HST} images, is $\sim9.2\times1.0$~kpc$^2$, which is roughly
consistent with the size of the SMC.

{The measured Vega-magnitudes are ${22.383\pm0.015}$ mag in F475W and ${22.470\pm0.018}$ mag in F625W.}
The F475W/F625W filters are similar (but not identical) to the Sloan $g/r$ filters used for the P60 telescope, 
with transformations typically being of order 0.1 mag.
These observations were taken at a phase of 1185~d from discovery (which corresponds to 1146~d in the rest frame).
A small part of the uncertainty arises from the background sky level, but the reported photometric uncertainty is dominated by Poisson noise from the object. 
The transformation by the exact filter profile as compared to the strong H$\alpha$ emission line may, however, be a greater concern.

\subsection{\textit{Swift}}
\label{sec:swift}
 
A first epoch of \textit{Swift} data was taken in May 2015 and  resulted in x-ray upper limits already discussed by A17.  \textit{Swift} 
was further triggered on seven additional occasions between June 2017 and February 2018; see Table~\ref{tab:swift}\footnote{Swifts PIs: 2015 Arcavi; 2017 Taddia, Rubin; 2018 Terreran.}. 
The epochs of \textit{Swift} observations are also indicated in Fig.~\ref{lc}.

Here, we discuss the observations at the location of iPTF14hls with the onboard X-Ray
Telescope (XRT; \citealt{burrows2005}).  Following A17, we use
online analysis tools \citep{evans2009} to search for X-ray emission at the location of iPTF14hls. 

We did not detect any source. The 90\% upper limit
on the $0.3$--$10.0$~keV count rate for each epoch is reported in Table~\ref{tab:swift}.
Combining the three latest, almost contemporary, epochs taken on February 2018 amounts to a total XRT exposure time of 11,855\,s (3.3\,hr), and an upper limit of
$4.9 \times 10^{-4}$ counts per second. 
If we assume a power-law spectrum with a photon index of $\Gamma
= 2$ and a Galactic hydrogen column density of $1.4\times10^{20}$
cm$^{-2}$,
this would correspond to an upper limit on the unabsorbed $0.3$--$10.0$
keV flux of $1.8 \times 10^{-14}$~erg~cm$^{-2}$~s$^{-1}$.
At the luminosity distance of iPTF14hls this corresponds to a
luminosity limit of $L_{\rm X} < 5.3 \times 10^{40}$
~erg~s$^{-1}$, at an epoch of $\sim 1194$ rest-frame days since discovery.
The X-ray limits are further discussed in Sect.~\ref{sec:xrays}. 

\section{Light curves}
\label{sec:lc}
\begin{figure}
\includegraphics[width=18cm]{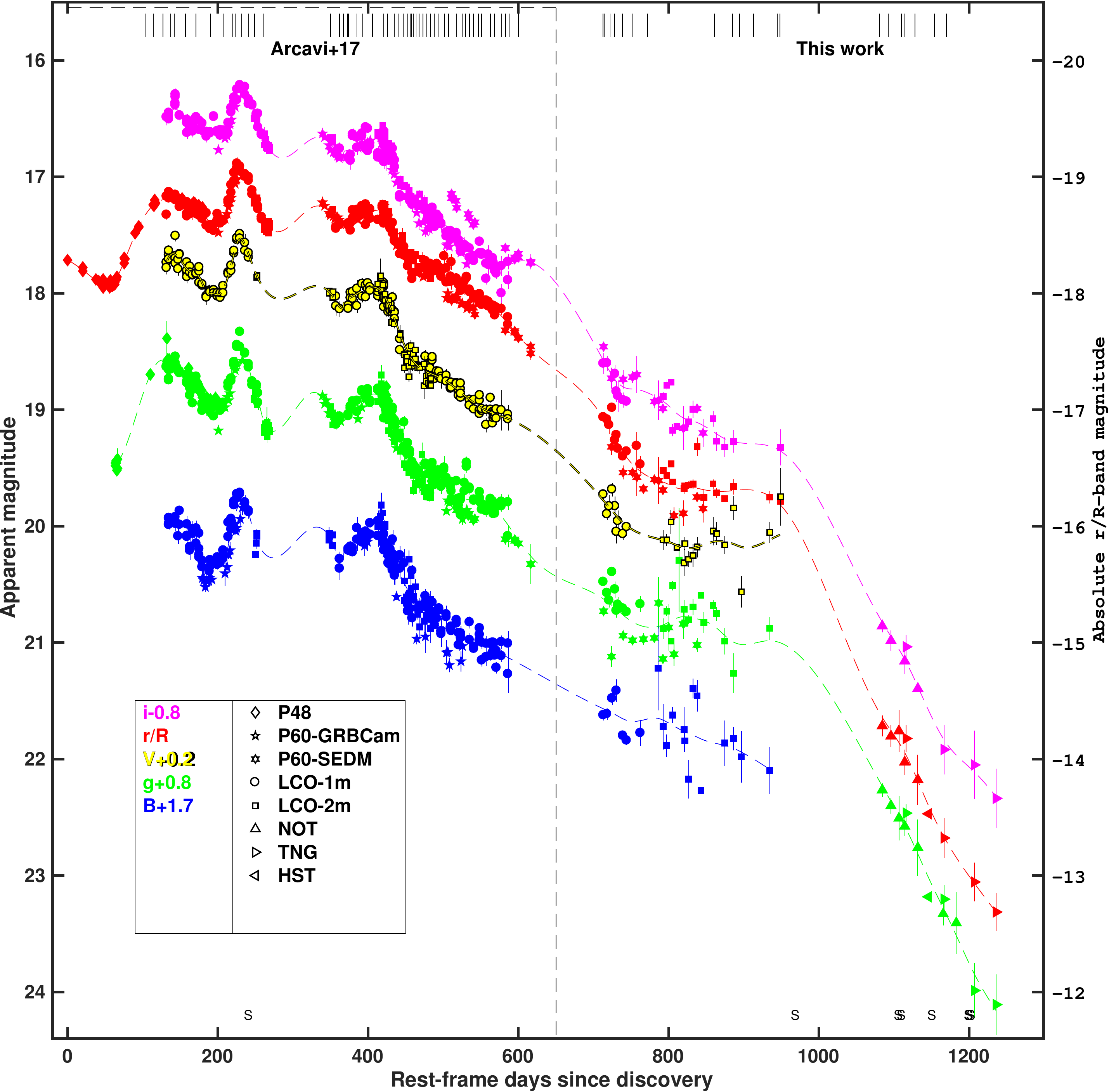}
\caption{\label{lc}Photometric observations of iPTF14hls. 
The data from the first 600~d, to the left of the vertical dashed line, are from A17. 
We have used the same telescopes for the continued monitoring, as well as NOT, TNG, and {\it HST} for the latest epochs. The vertical bars on top of the plot indicate epochs of spectroscopic observations, and the ``S" symbols in the bottom part indicate the epochs of \textit{Swift} observations. Each light curve was fitted with a combination of two tension splines (dashed coloured lines) to provide a continuous representation of the data. On the right-hand ordinate we report the absolute magnitude for the $r/R$ band. To obtain the absolute magnitudes for the $BgVi$ bands, their apparent magnitudes must be shifted by $-36.024$ mag, $-36.019$ mag, $-36.010$ mag, and $-35.993$ mag, respectively.
}
\end{figure}

\begin{figure}
\includegraphics[width=9cm]{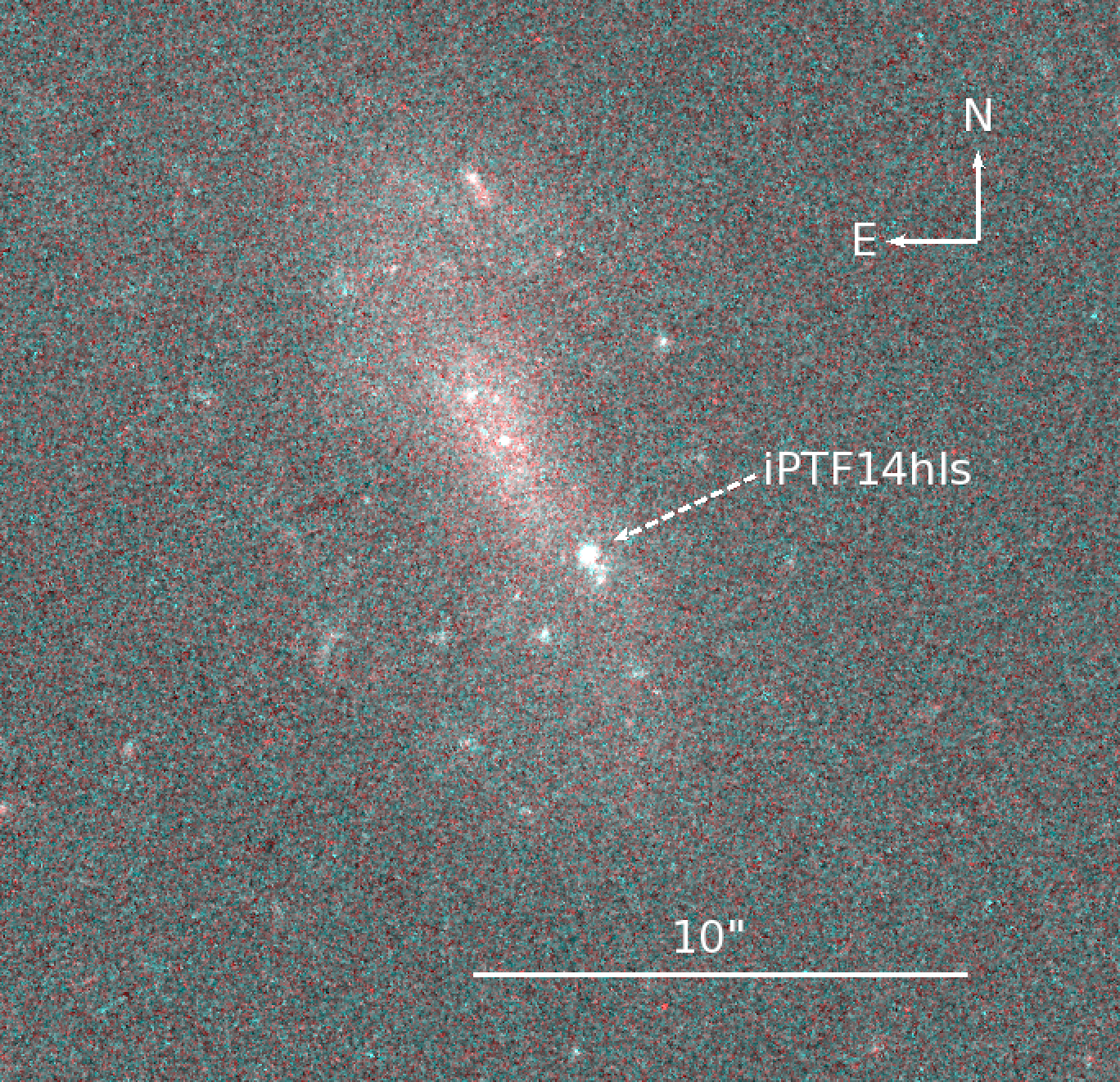}
\caption{\label{HST} {\it HST} F475W and F625W composite image. The {\it HST} images of iPTF14hls and its host galaxy were obtained using the WFC3 on 20 December 2017, or 1185 rest-frame days past discovery. One arcsecond corresponds to 720 pc. We have indicated the SN itself with an arrow, which is clearly seen. Also, a nearby \ion{H}{ii} region can be discerned.}
\end{figure}

Figure~\ref{lc} shows all the optical LCs, fit by tension splines to guide the eye and give a continuous representation of the data. A17 obtained their last photometric epochs at around 615~rest-frame days (the rest frame is used from now on unless otherwise specified), 
when iPTF14hls had an $r$-band magnitude of $\sim18.5$. 
A decline occurred in the following months while the SN was behind the Sun. We recovered iPTF14hls at $713$~d at $r \approx 19.1$~mag, which corresponds to a decline of 0.6 mag in 100~d. The SN was also recovered in the $B$, 
$g$, $V$, and $i$ bands, with similar drops in brightness (Fig.~\ref{lc}).  

The $V$, $r$, and $i$ bands then continued to decline 
until about 800~d, whereafter they settled on a fairly constant value ($V$ and $r$) or declined more slowly ($i$~band). A similar behaviour is shown by the $B$ and $g$ LCs, although these LCs are more scattered.
After about 950~d, follow-up observations were interrupted for $\sim100$~d, and then all the LCs ($g$, $r$, $i$) were recovered about 
1.5 mag below the last detection. All bands then show a faster decline between 1084 and 1236~d, when they drop by another $\sim 1.5$--2.0 mag. The slope of the $r$-band LC is 1.15 mag (100 d)$^{-1}$ in this period.

We also report the absolute $r/R$-band magnitudes of iPTF14hls on the right-hand side of Fig.~\ref{lc}.
The SN, which peaked at $r = -19.1$~mag and stayed brighter than $-18$ mag for a bit more than 500~d, is found to lie between $M_r = -17$ and $-16$~mag between 700 and 950~d past discovery. 
At the latest epochs after 1080~d, 
$M_r$ ranges between $-14.5$~mag and $-12.7$~mag. 

\begin{figure}
\includegraphics[width=18cm]{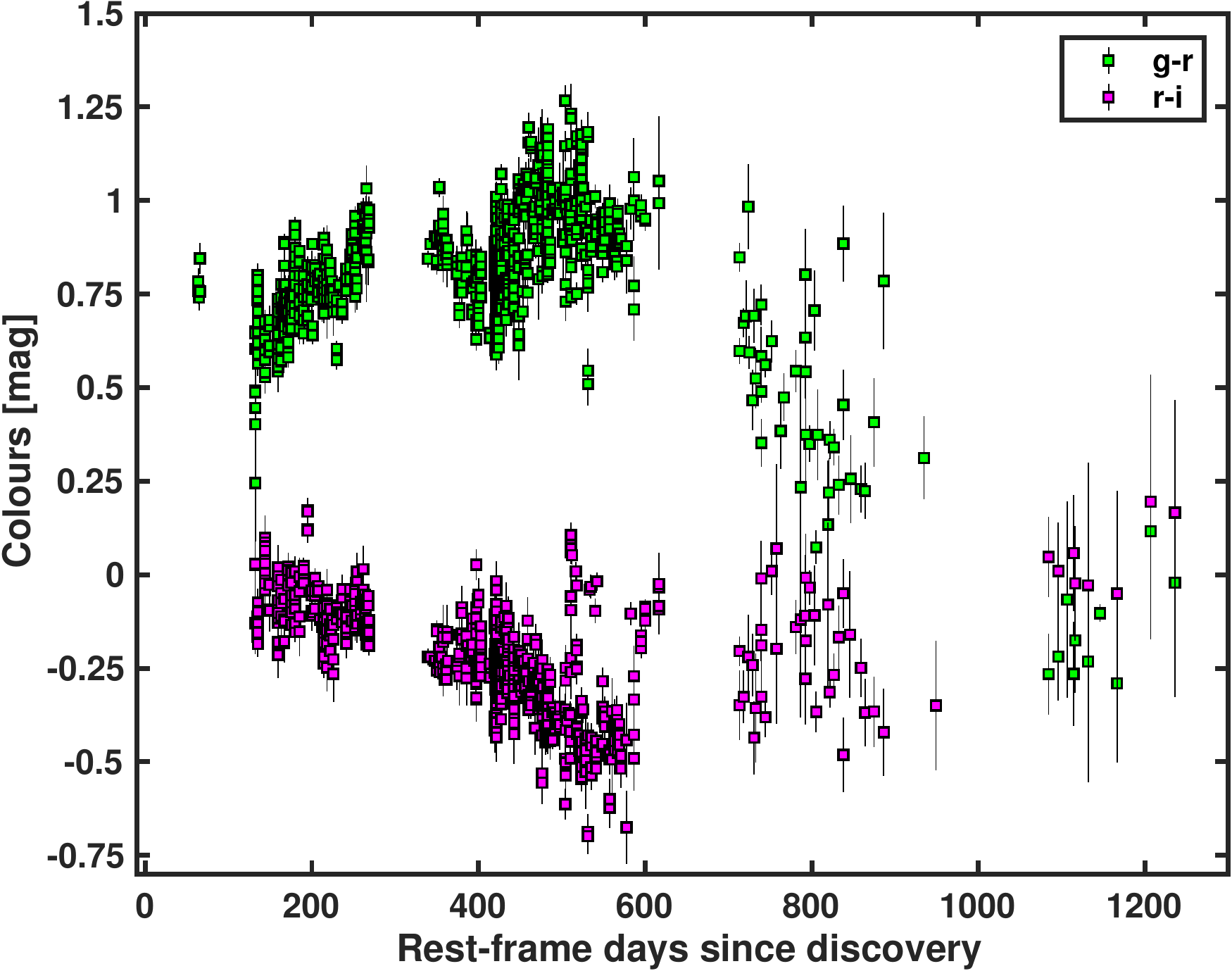}
\caption{\label{color} Colour evolution of iPTF14hls in both $g-r$ and $r-i$, where the data from the first 600~d are from A17.}
\end{figure}

The colour evolution of iPTF14hls is shown in Fig.~\ref{color}. 
During the first 600~d, $g-r$ slowly evolves toward redder values. 
After this date the trend is reversed and the colour initiates a quick evolution toward bluer values, reaching $\sim -0.25$ mag.
After 1000~d it may again evolve to slightly redder colours. 
The $r-i$ colour moves bluewards over the first 600~d, and thereafter moves slowly toward the red (Fig.~\ref{color}).

\begin{figure}
\includegraphics[width=18cm]{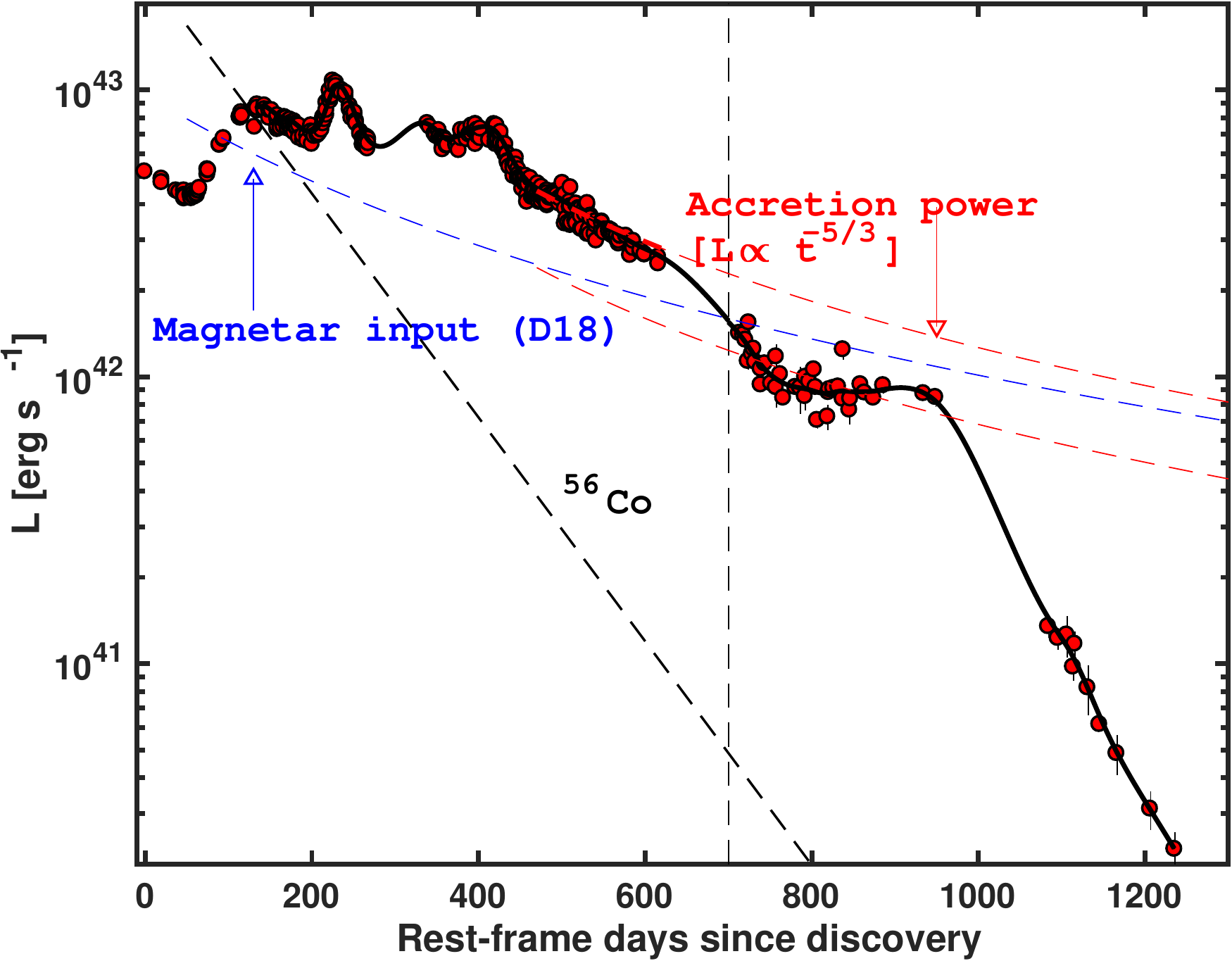}
\caption{\label{bolo}iPTF14hls bolometric LC. Before 700~d, we use the BB fit to the SEDs to compute the luminosity. After 700~d, when the spectra can no longer be approximated by BBs, we use bolometric corrections obtained from the spectra, which we add to the $r$ band to compute the pseudo-bolometric LC. 
The solid black line is the bolometric LC obtained from the spline-interpolated photometry --- that is, from the dashed curves in Fig.~\ref{lc}. To highlight the scatter in the actual observed photometry, which dominates the uncertainty in the bolometric LC, we also apply the bolometric corrections to the observed individual $r$-band data points, and show the result with red circles.
We assume that at early epochs, when only the $r$ band was observed, the bolometric correction is constant and equal to the one computed from the BB fit to the SED at 133 rest-frame days, when the $g$ and $i$ bands also were observed. For epochs later than 1170~d, we assume the bolometric correction is constant and equal to that computed from the last spectrum. 
We also show a PL (dashed red line) with $n=-5/3$ fitting the epochs between 470 and 620~d. The PL does not reproduce the luminosity at later epochs. If we instead scale the PL to fit the epochs between 700 and 950~d, it is still not in accordance with the fast decline after 950~d. We also report the $^{56}$Co power input (black dashed line), which would require 
an unfeasible amount of
nickel to power the +950~d decline. The magnetar energy input (blue dashed line) from \citet{Dessart18} roughly reproduces the LC decline until 950~d, but not the following sharp decline.}
\end{figure}

Using the multi-band LCs and the spectra to estimate bolometric corrections, it is possible to compute the bolometric LC of iPTF14hls until late epochs. 
First we converted the $gri$ magnitudes into specific fluxes at their effective wavelengths, including the correction for extinction. This spectral energy distribution (SED) at each epoch can be fit with a black-body (BB) function, as in A17. The integral of the BB fit to the SEDs times $4\pi D^{2}$, where $D$ is the luminosity distance, gives the SN luminosity (see Fig.~\ref{bolo}). A BB is a reasonable match to the SED until $\sim 700$~d, after which the SED deviates  
from a BB as the emission lines dominate the nebular spectra (see Sect.~\ref{sec:spectra}). Until 700~d, the BB fits give an average temperature of $5360 \pm 250$~K, consistent with the H-recombination temperature, as noted by A17. 

To construct the bolometric LC after 700~d, we have to take into account the exact shape of the SED. We computed bolometric corrections for the $r$-band magnitudes using the spectral sequence presented in Sect.~\ref{sec:spectra}. We have first obtained an absolute calibration of these spectra by computing the synthetic $r$-band photometry from the spectra and then scaling the spectra to the measured template-subtracted $r$-band photometry. 
We then integrated the spectra that cover the range between 4000 and 9200~\AA\ to get the integrated flux in this optical range. Next, we multiplied this flux by $4\pi D^{2}$ to obtain the luminosity. We converted the luminosity into a pseudo-bolometric magnitude, and compute the bolometric corrections as the difference between these pseudo-bolometric magnitudes and the observed $r$-band magnitudes at the epoch of each spectrum. We fitted these bolometric corrections with a low-order polynomial between 700~d and 1240~d, and then apply the resulting corrections to the entire late-time $r$-band LC to obtain the pseudo-bolometric LC between 700 and 1240~d. This is shown in Fig.~\ref{bolo}. This pseudo-bolometric LC from the optical spectra therefore does not include the emission in the ultraviolet (UV) or in the near-infrared (NIR). However, it matches the bolometric LC at 700~d as obtained from the BB fit, where the UV and NIR flux is included. We thus assumed no further corrections to the pseudo-bolometric LC after 700~d, assuming that most of the flux is (still) in the optical.

The bolometric LC computed with the outlined method is shown by a thick solid black line in Fig.~\ref{bolo}. This comes from the photometry obtained from the interpolation of the spline fits shown in Fig.~\ref{lc}. In order to better illustrate the actual scatter in the observed photometry, which dominates the statistical uncertainty in the bolometric LC, we also overplot (red circles) the bolometric LC as derived from the individual observed $r$-band points to which we applied the bolometric corrections described above (i.e. based on the BB luminosity before 700 d, and on the spectra afterward). The bolometric LC from the $r$ band {is available on WISeReP}.

\section{Spectra}
\label{sec:spectra}

A17 presented a comprehensive dataset of low-resolution optical spectra of iPTF14hls, in total 62 spectra for the first 600~d. They revealed normal SN~II spectra, but with an unprecedented slow evolution.

We continued the spectroscopic monitoring and here present an additional 
23
spectra covering the period from 713 to 1170 rest-frame days past discovery. These spectra were obtained  with a suite of different telescopes and instruments. 

Eight of the spectra were obtained with ALFOSC mounted on NOT, whereas the LCO Haleakala (FTN) telescope was equipped with the low-dispersion spectrograph FLOYDS and provided five spectra.
Two spectra were obtained with DOLORES mounted on the TNG.
From Hawaii we also obtained
four
spectra with Keck I, equipped with the Low Resolution Imaging Spectrometer (LRIS; \citealp{oke95}), and four Keck II spectra using DEIMOS. 
The log of the spectral observations is given in Table~\ref{tab:spectra}. All spectra {are publicly available} at WISeREP \citep{wiserep}. Spectral reductions were carried out with standard procedures, including wavelength calibration via arc-lamp spectra, and flux calibration with spectrophotometric standard stars. 

\begin{figure}
\includegraphics[width=18cm]{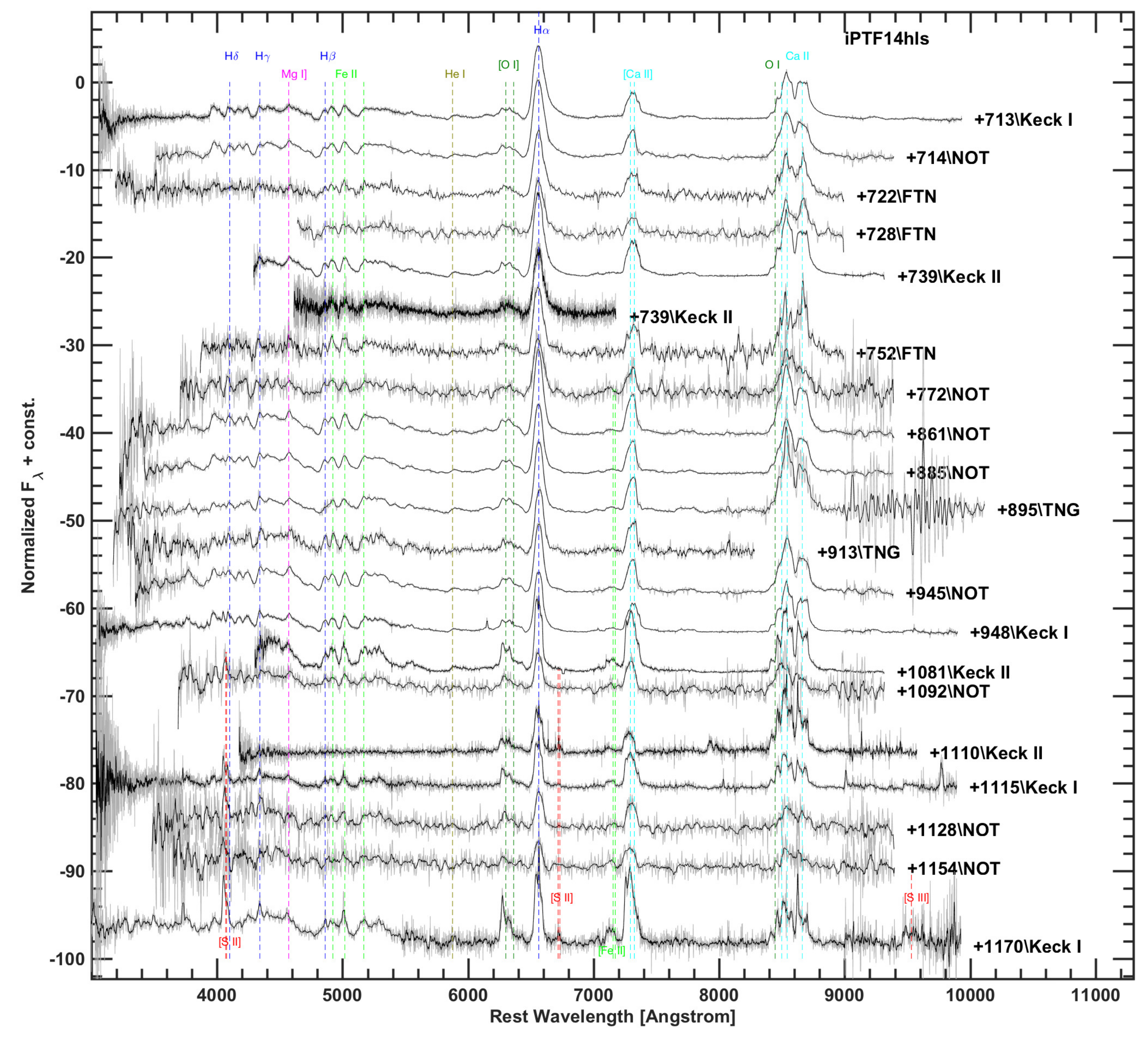}
\caption{\label{specseq}Late-time ($>700$~d) spectral sequence of iPTF14hls. The SN went into the fully nebular phase during this period. Some of the main features are labelled and marked by coloured dashed lines at their rest wavelengths. The rest-frame phases are indicated next to the spectra. The spectra have been offset in flux for clarity. 
}
\end{figure}

The main features of these late-time spectra are still compatible with those of a SN~II (see Fig.~\ref{specseq} for line identifications).
The most prominent line is H$\alpha$. 
As the SN evolves, the [\ion{Ca}{ii}]~$\lambda\lambda$7293, 7324 doublet becomes relatively brighter, almost reaching the strength of H$\alpha$ at 1170~d. All the spectra are characterised by the presence of the NIR \ion{Ca}{ii} triplet, and weak \ion{O}{i}~$\lambda$8446 next to it. The [\ion{O}{i}]~$\lambda\lambda$6300, 6364 doublet is observed in emission and it is particularly strong in the last spectrum, where we also
identify a bright and not so common [\ion{S}{ii}]~$\lambda\lambda$4069, 4076 emission line dominating the bluer part of the spectrum. The corresponding [\ion{S}{ii}]~$\lambda\lambda$6716, 6731 doublet is barely detected in the same spectrum. Closer inspection reveals that the [\ion{S}{ii}]~$\lambda\lambda$4069, 4076 doublet starts emerging when H$\alpha$ becomes asymmetric, after 948~d. The [\ion{S}{ii}]~$\lambda\lambda$4069, 4076 emission line is further discussed and analysed in Sect.~\ref{sec:PPIS}.

Clear P-Cygni features associated with \ion{Fe}{ii}~$\lambda\lambda$5169, 5018, 4923, as well as H$\beta$, H$\gamma$, and H$\delta$, characterise the blue part of the spectra. There is also emission of \ion{Mg}{i}]~$\lambda$4571 at all epochs. 
At epochs later than 770~d, [\ion{Fe}{ii}]~$\lambda$7155 emerges. 

\begin{figure}
\includegraphics[width=9cm]{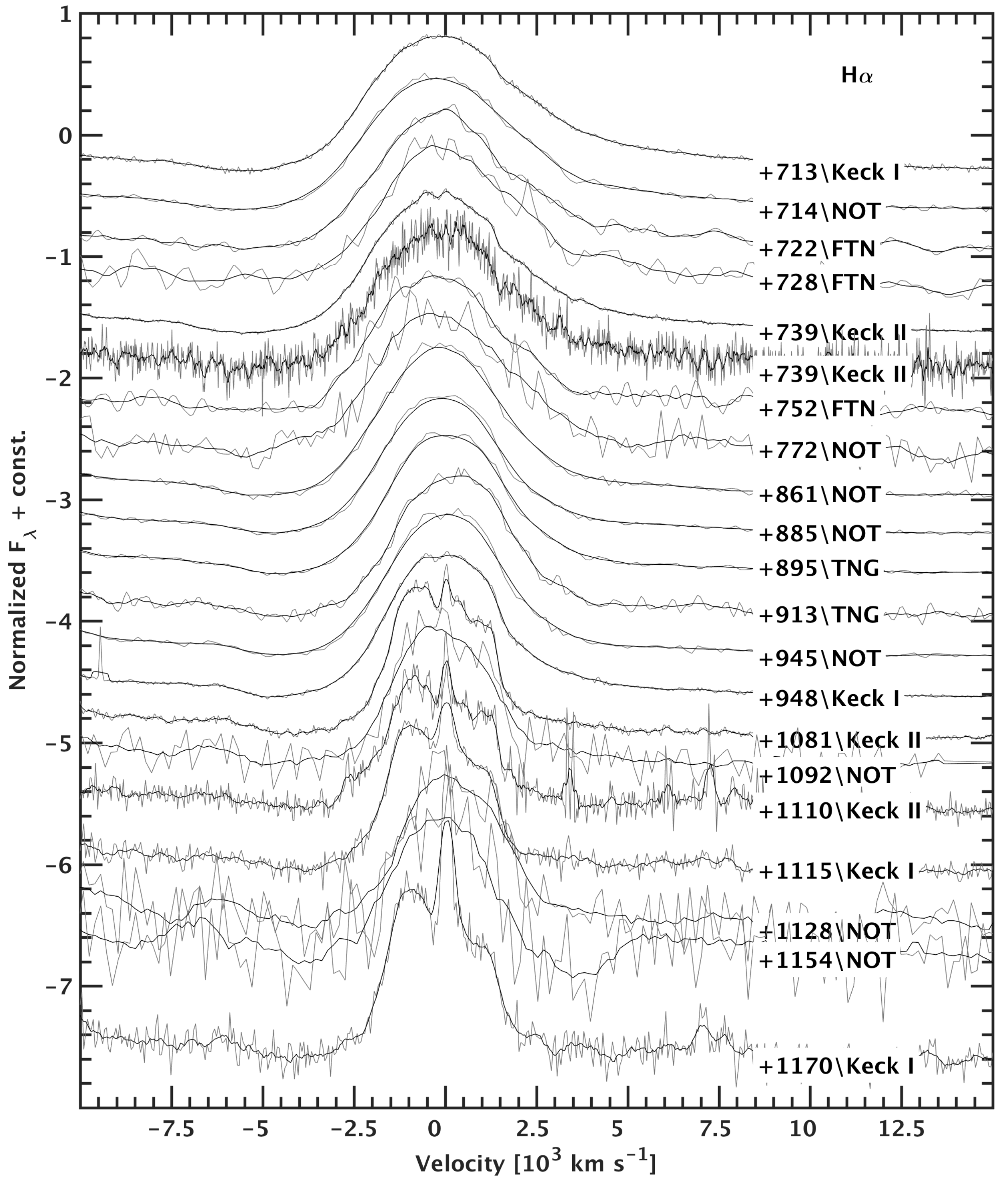}
\includegraphics[width=9cm]{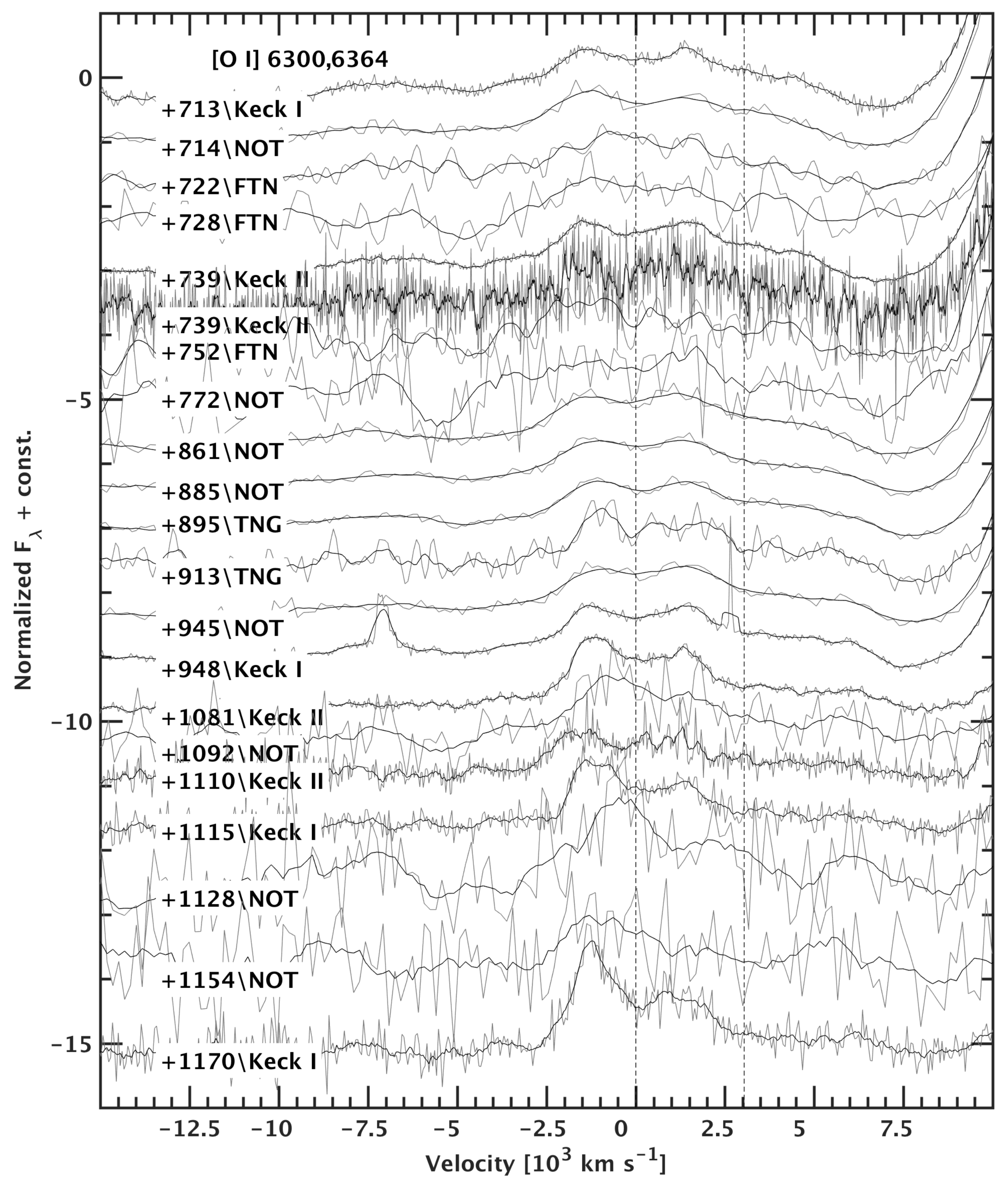}
\caption{\label{HaOI} (Left-hand panel) H$\alpha$ line profiles from the spectra of iPTF14hls. (Right-hand panel) [\ion{O}{i}]~$\lambda\lambda$6300, 6364 from the spectra of iPTF14hls. The 
black dashed lines mark the
position of the 6300 and of the 6364 components, assuming $\lambda = 6300$~\AA\ at zero velocity. 
}
\end{figure}

\begin{figure}
\includegraphics[width=9cm]{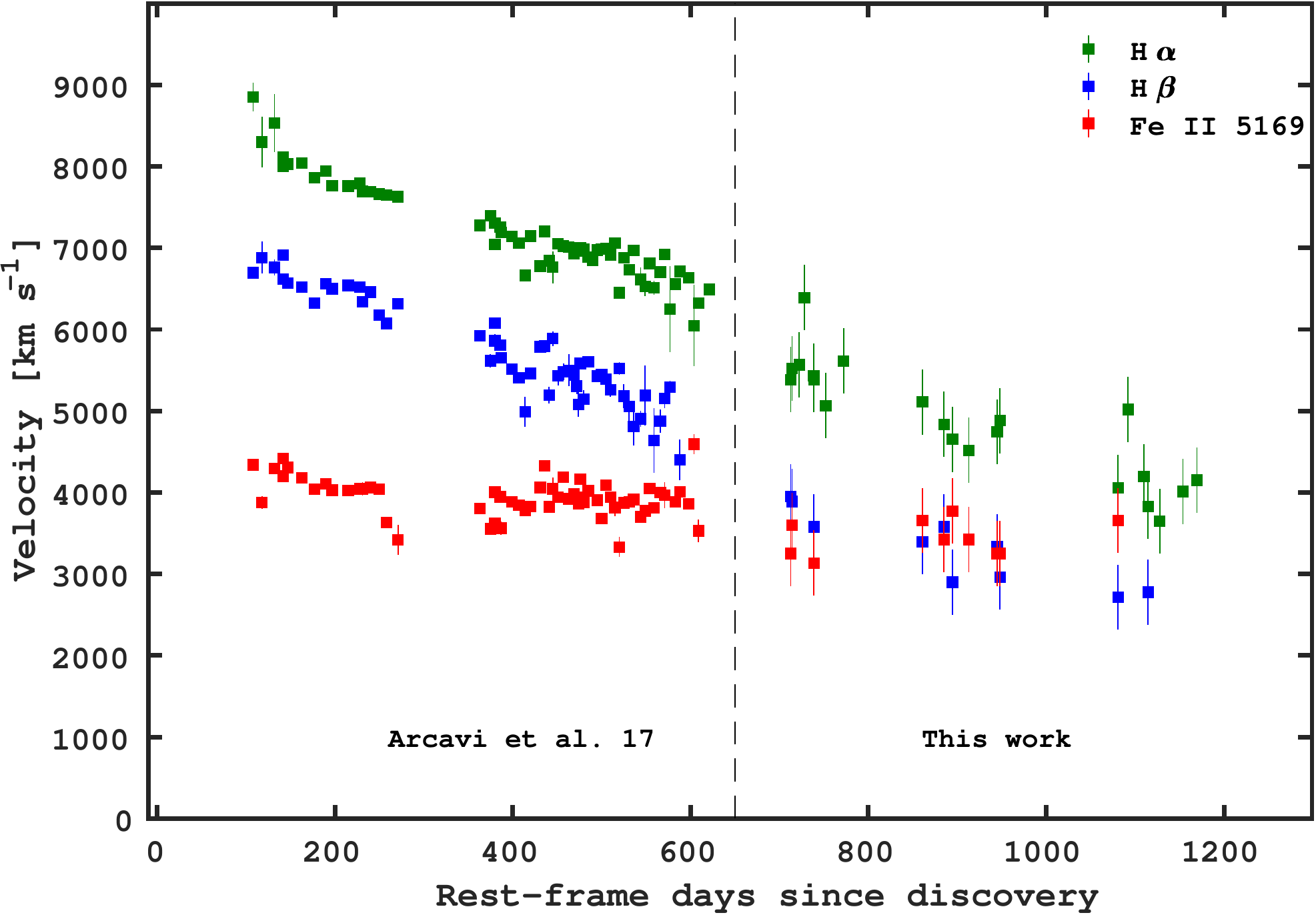}
\caption{\label{vel} H$\alpha$, H$\beta$, and \ion{Fe}{ii}~$\lambda$5169 velocities as measured from the spectra of iPTF14hls. Left of the vertical dashed line displays measurements for epochs presented by A17, while the late-time velocities from our new spectra are displayed to the right of this line. The Fe line velocity remains constant in time, whereas the Balmer-line velocities continue to decline slowly.}
\end{figure}

The evolution of the H$\alpha$ profile is shown in closer detail in Fig.~\ref{HaOI} (left-hand panel). Until 948~d, the line exhibits a P-Cygni profile whose emission part is clearly more pronounced.
The absorption minimum of H$\alpha$, which becomes hard to identify at epochs later than 948~d, seems to move to lower velocities with time (see Fig.~\ref{vel}), from about 5500~km~s$^{-1}$ at $\sim 700$~d down to 4000 km~s$^{-1}$ at $\sim 1200$~d (where it is more an indication of the blue velocity at zero intensity 
of the emission component).
The H$\alpha$ emission is rather symmetric for the first 948~d, 
and there is not much evolution in the line profile despite the 200~d time interval. At 1081~d, the emission profile becomes asymmetric (Fig.~\ref{HaOI}, left-hand panel), with the red side suppressed compared to the blue side of the emission line. There is a narrow (unresolved) emission line centered at zero velocity that is seen in the spectra with higher resolution (Keck I and II). This narrow line was also reported by \citet{AS17} at +1153~d, and we discuss the origin of this line in Sect.~\ref{sec:narrow}.

The emerging asymmetry of H$\alpha$ at epochs later than $\sim1000$~d seems to be similar to the one showed by [\ion{O}{i}]~$\lambda\lambda$6300, 6364, where the blue component is also stronger than the red one at later phases, whereas at earlier epochs they seem to be equivalent in strength (see Fig.~\ref{HaOI}, right-hand panel). 

\begin{figure}
\includegraphics[width=9cm]{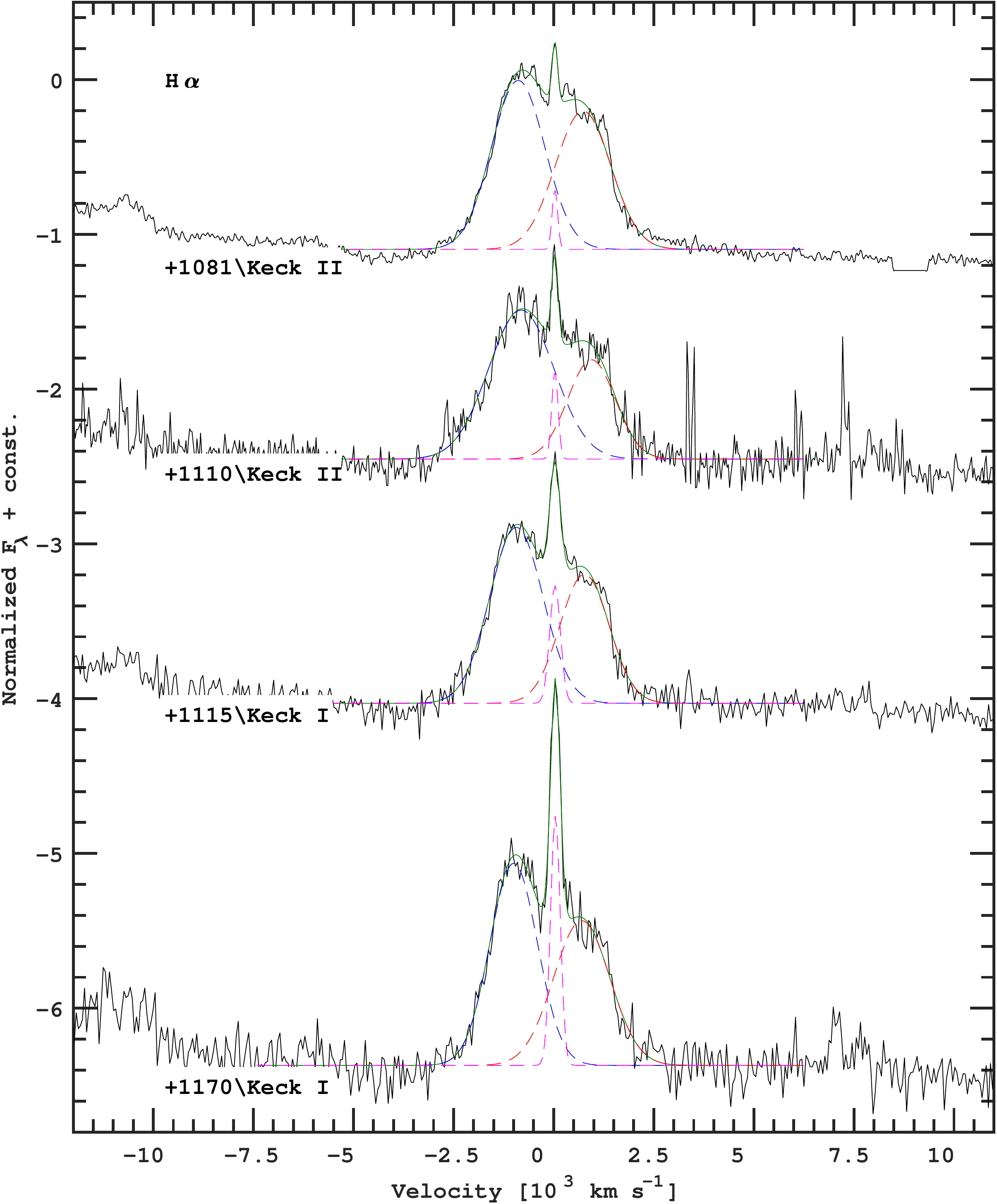}
\includegraphics[width=9cm]{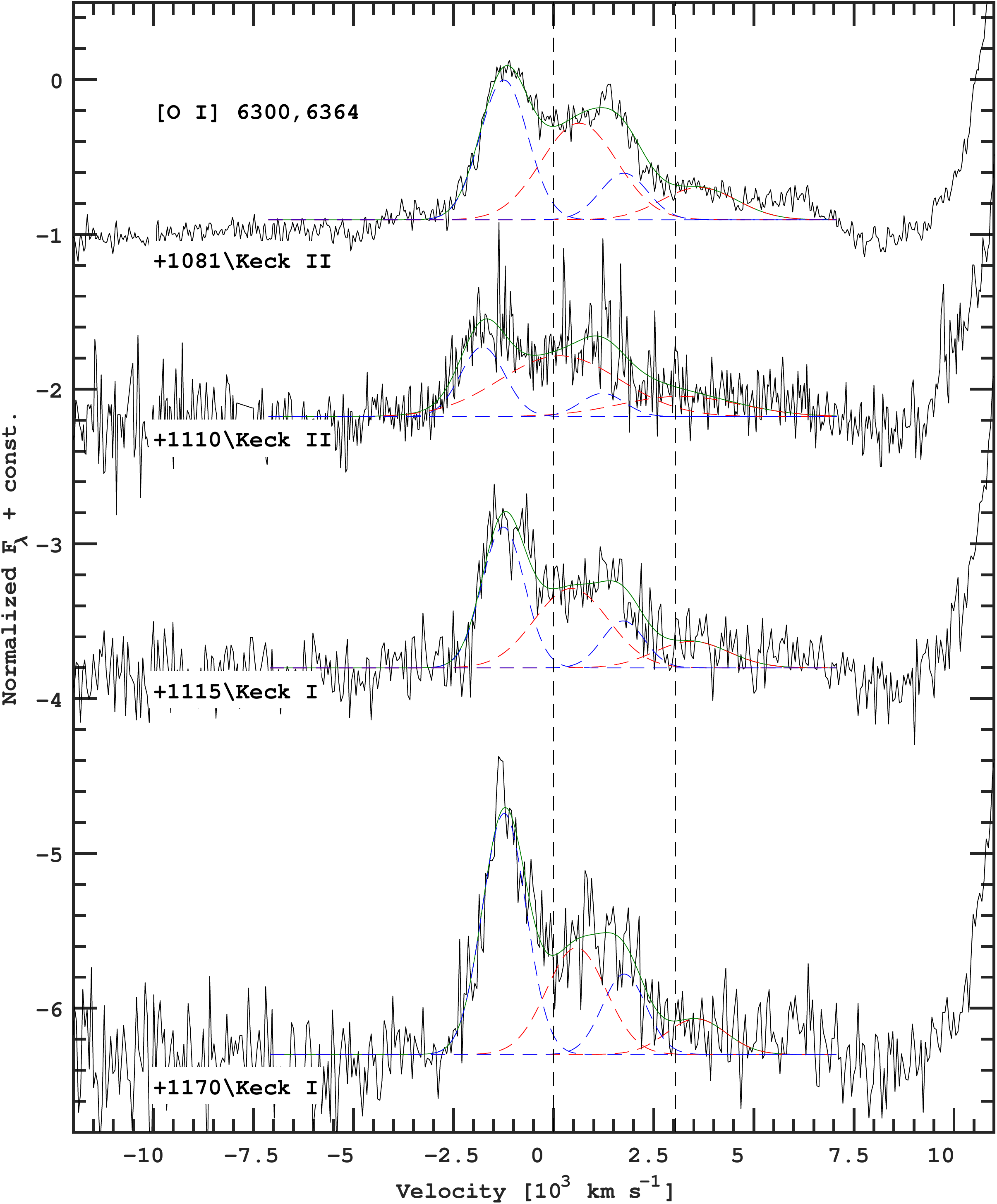}
\caption{\label{HaOI3fit}As in Fig.~\ref{HaOI}, but here the four highest signal-to-noise ratio late-time spectra are plotted around the wavelength range of H$\alpha$ (left) and [\ion{O}{i}]~$\lambda\lambda$6300, 6364 (right). For H$\alpha$, a triple Gaussian fit (green line for the total fit; blue, red, and magenta for the single components) is performed to discuss the late-time asymmetry of the line. For  [\ion{O}{i}]~$\lambda\lambda$6300, 6364, we  used two Gaussians (a blueshifted and a redshifted component) for each of the two lines (6300 and 6364~\AA). The normalisation of the two lines was fixed so that the 6300~\AA\ line has three times the flux of the 6364~\AA~line. The centroids of the corresponding components of the two lines were fixed to be 64~\AA\ apart.}
\end{figure}

The late-time ($>1000$~d) H$\alpha$ and [\ion{O}{i}]~$\lambda\lambda$6300, 6364 profiles can be fit by a combination of two (three counting the narrow component in the case of H$\alpha$) Gaussians as shown by \citet{AS17}, 
one for the blue and one for the red side (see Fig.~\ref{HaOI3fit}, where we fit this model
to the three Keck spectra). The same is true for the [\ion{S}{ii}]~$\lambda\lambda$4069, 4076 line.  
In the case of the [\ion{O}{i}]~$\lambda\lambda$6300, 6364 blend, in order to properly reproduce the profile we  
add an extra double-Gaussian component to fit the 6364~\AA\ line on the red side of the 6300~\AA\ line (fixing their ratio to 1:3).

\begin{figure}
\includegraphics[width=12cm]{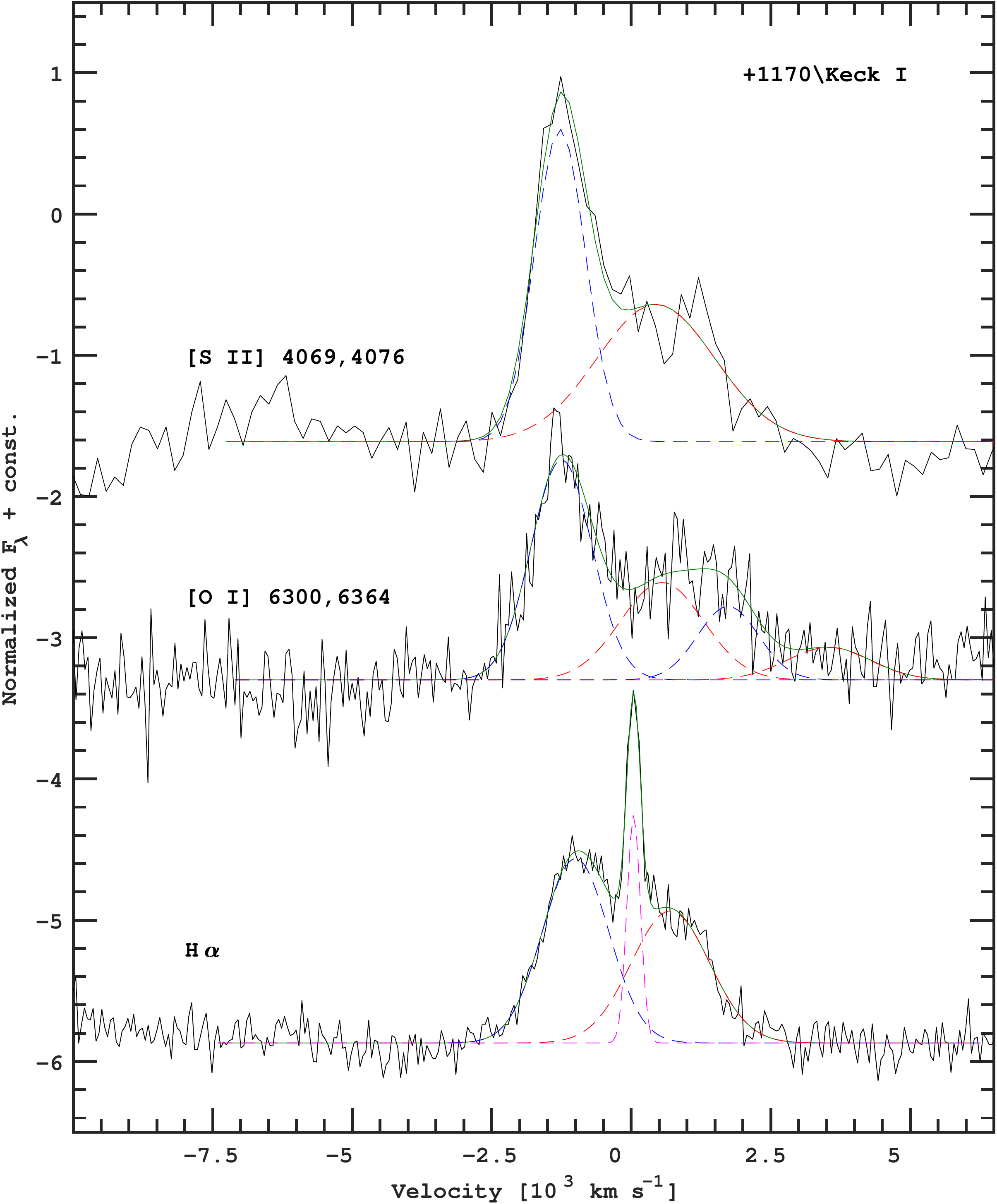}
\caption{\label{lineprofiles}
Gaussian fits to three strong emission lines in the last Keck spectrum.
The [\ion{S}{ii}]~$\lambda\lambda$4069, 4076,  [\ion{O}{i}]~$\lambda\lambda$6300, 6364 and H$\alpha$ profiles are here shown in common velocity space (the zero velocities correspond to 4068.6~\AA, 6300.3~\AA, and 6562.8~\AA, respectively). 
The decompositions into Gaussians are tabulated in Table~\ref{tab:linecomp}.
}
\end{figure}

If we compare the shift with respect to zero velocity of the Gaussian components that reproduce these lines, and their full width at half-maximum intensity (FWHM; Table~\ref{tab:linecomp}), it seems clear that all three lines are produced at a similar location in the ejecta. To further illustrate this we plot in Fig.~\ref{lineprofiles} these three emission lines in velocity space. 
The uncertainties in Table~\ref{tab:linecomp} are statistical errors on the fit, estimated as the standard deviation of 200 Monte Carlo realisations 
where the noise of the spectrum was taken into account.

Concerning the different lines of the \ion{Ca}{ii} NIR triplet and in particular the 8662~\AA\ feature, they also seem to show an increasing asymmetry at similar epochs, with the bluer part dominating.  
The [\ion{Ca}{ii}]~$\lambda\lambda$7293, 7324 blend shows the same asymmetry only in the last spectrum.

\section{Discussion}  
\label{sec:discussion}

Although several papers quickly came up with a suite of possible powering mechanisms for iPTF14hls, few of these made predictions for the post-600~d evolution.
It is therefore not our intention to make detailed comparisons with the published models, but we do offer a few remarks.

\subsection{Accretion}
\cite{WangWangWang18} put forward a model with erratic fallback accretion onto a compact object to explain the long-lived bumpy LC. With a large number of parameters for their accretion model (and adding a magnetic eruption) they can indeed reproduce the observed LC
from A17. However, given that accretion decreases with time, their model should predict a late-time luminosity declining as a 
power law (PL) with index $n$ ($L \propto t^n$), where $n = -5/3$. Such a decline was already 
suggested to reproduce the emission after the last bump in the bolometric LC by A17.
In Fig.~\ref{bolo} we show that at later epochs this may hold
until about 620~d, when the bolometric LC starts declining faster than $n = -5/3$.  Alternatively (Fig.~\ref{bolo}), the late-time emission between 600 and 900~d could be due to accretion. However, the very late-time ($> 1000$~d) LC is not consistent with accretion. In fact, 
the best-fit PL to the post-950~d bolometric LC has an index of $n = -13.5$. 
{The late time LC therefore do not favour the accretion scenario.}

\subsection{Radioactivity}

The radioactive powering scenario --- common for SNe~II at late phases --- was already ruled out by (for example) A17 and \cite{Woosley18}. 
For completeness we mention it here, since 
the very late-time LC slope is not so different from the decay rate of 
$^{56}$Co; the LC slope at 1000--1200 days is 1.4 mag per 100 days (versus 0.098 mag (100 d)$^{-1}$ for radioactivity). However, we note here that the actual mass of radioactive $^{56}$Ni that needs to be ejected in the explosion to power that emission at 900--1200 days is $\sim140~M_\odot$. This could be lowered somewhat by adding $^{57}$Co and $^{44}$Ti \citep[e.g.][]{fransson1993, sollerman2002}, but is
anyway clearly an unfeasible suggestion, that would also have made the SN much too bright at peak. 
{This powering mechanism can thus be ruled out.}

\subsection{Magnetar}
The magnetar model was not favoured by A17 given the high model luminosity at early times required to reproduce the late-time decline. Including diffusion, \citet{Dessart18} mediates this problem and even gets a too dim early-time LC, which could likely be resolved simply by using a more extended star. The \citet{Dessart18} magnetar model for iPTF14hls
works well in terms of reproducing the spectral evolution the first 600~d. 
In their model a4pm1, the electron-scattering optical depth scales like $\tau_{\rm es} \propto t^{-2}$ and is about 1.3 at 600~d. Therefore, the ejecta are still optically thick at 600~d, and H$\alpha$ will remain optically thick even longer. We observed H$\alpha$ becoming narrower and asymmetric after 948~d 
(Sect.~\ref{sec:spectra}), 
when the ejecta are likely optically thin, which is not inconsistent with this scenario.

The magnetar model of \cite{Dessart18} could also well provide the luminosity for the time span 100--600~d, although not necessarily the undulations. Such undulations are often seen in SNe~IIn \cite[e.g.][]{nyholm}. There are, however, no explicit predictions by \cite{Dessart18} about the post-600~d evolution of iPTF14hls. 

In Fig.~\ref{bolo} we include an extrapolation of the magnetar model proposed by \citet{Dessart18}, which clearly does not fall off as observed after 1000~d. Such a steep decline
would require either an {ad hoc} switchoff of the magnetar or at least a very rapid drop in the energy deposition. 
Perhaps one could envision ways in which a magnetar-powering mechanism undergoes dramatic changes, for example if the bubble breaks out of the ejecta \citep{blondin},
but we emphasise that there is also little change in the late-time spectra. A dramatic breakout, affecting the LC, should likely also be accompanied by a change in the spectral appearance, with more high-ionisation lines and also probably a higher X-ray luminosity \citep{metzger}.
This is not seen in our data.

\subsection{CSM interaction: Pulsational pair-instability supernova}
\label{sec:PPIS}

We estimate the total energy emitted by iPTF14hls between discovery and the last observation (1235~d) to be $3.59 \times 10^{50}$~erg, 
which is $\sim 90\%$ of the energy available in the above-mentioned magnetar model and also easily accommodated by CSM models discussed below. It is $\sim 70$\% of the energy available for non-rotating pulsational pair-instability supernova (PPISN) models suggested by \cite{Woosley2017}, so in terms of radiation energetics this long-lived SN allows for several explanations.

One potential key feature to rule out a PPISN shell-shell interaction scenario - which is a special case of a more general CSM interaction model (Sect.~\ref{sec:csm}) 
- is to search for signs of core collapse, such as
evidence for central nucleosynthesis, which is not predicted in such a PPISN scenario.

Our latest spectrum shows a strong emission line of [\ion{S}{ii}]~$\lambda\lambda$4069, 4076  that is not always present in late-time nebular SN spectra. 
This line was detected in SN~1980K almost 15~yr past explosion \citep{fesen}, and 
the detection of 
[\ion{S}{ii}]~$\lambda\lambda$4069, 4076  
emission but a lack of 
[\ion{S}{ii}]~$\lambda\lambda$6716, 6731  
was there suggested to imply an electron density of $n_e > 10^5$~cm$^{-3}$.

We do the same exercise here and measure the ratio of 
flux of [\ion{S}{ii}]~$\lambda\lambda$4069, 4076 and that of [\ion{S}{ii}]~$\lambda\lambda$6716, 6731 to be 
$\sim10$,
whereas 
the ratio of  flux of [\ion{O}{i}]~$\lambda$5577 and that of [\ion{O}{i}]~$\lambda\lambda$6300, 6364 is $<0.15$.

\begin{figure}
\includegraphics[width=12cm]{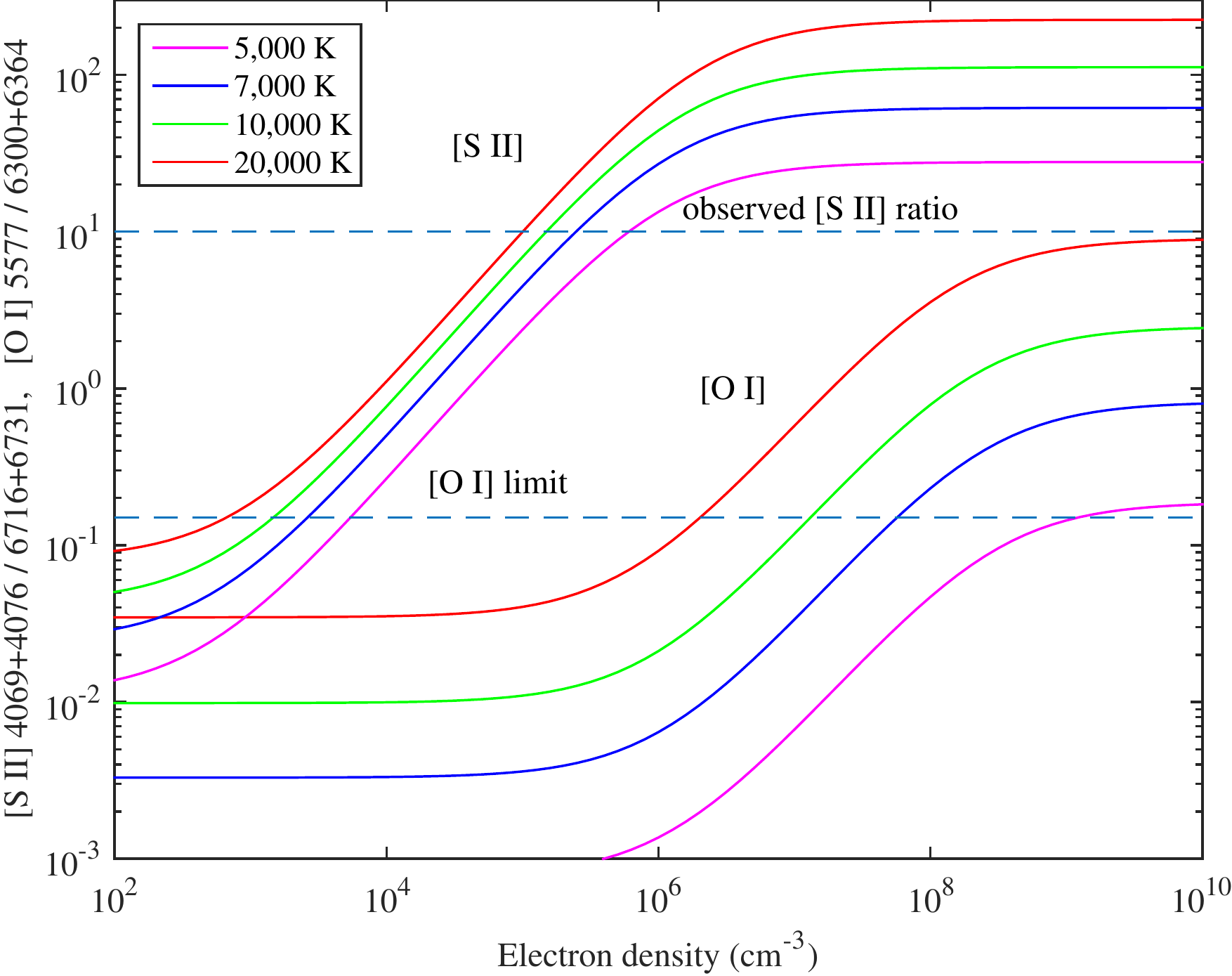}
\caption{\label{SII_OI_diag}
[\ion{S}{ii}]~$\lambda\lambda$4069, 4076 to [\ion{S}{ii}] $\lambda\lambda$6716, 6731 ratio as function of electron density for different temperatures, together  with the [\ion{O}{i}]~$\lambda$5577 to  $\lambda\lambda$6300, 6364 ratio. The observed [\ion{S}{ii}] ratio constrains the density to be (1.0--6.3) $\times 10^5$ cm$^{-3}$, consistent with the [\ion{O}{i}] limit.
}
\end{figure}

In Fig.~\ref{SII_OI_diag} we show the [\ion{S}{ii}]~$\lambda\lambda$4069, 4076 to [\ion{S}{ii}]~$\lambda\lambda$6716, 6731 ratio as a function of electron density for temperatures in the likely range 5000--20,000~K. The observed [\ion{S}{ii}] ratio of $\sim 10$ constrains the density to be (1.0--6.3) $\times 10^5$ cm$^{-3}$. At very high densities the corresponding auroral [\ion{O}{i}]~$\lambda$5577 line also becomes strong. We therefore plot  the [\ion{O}{i}]~$\lambda$5577 to [\ion{O}{i}]~$\lambda\lambda$6300, 6364 ratio in the same figure. The absence of this line is consistent with the above density range. Note that this refers to the electron density; if the ionisation fraction is low, as may be the case at these late epochs, the total density may be higher.

The emission-line analysis presented in Sect.~\ref{sec:spectra} assumed Gaussian profiles.  Following \cite{AS17}, we decomposed the nebular line profiles into two Gaussians, but also accounted for the doublet nature of the [\ion{O}{i}]~$\lambda\lambda$6300, 6364 line.

This exercise clarified that all the relevant elements emit from the same locations, and the typical velocity width of the lines suggests we are probing the central ejecta properties (Table~\ref{tab:linecomp} and Fig.~\ref{lineprofiles}). Of course, Gaussian profiles may not be the correct representation of the emission profile from the gas. Such a decomposition did seem to make sense (for example) for SN~1998S, which had an H$\alpha$ emission-line profile at late epochs that resembles what we see in iPTF14hls \citep{pozzo}.
However, we cannot rule out that the emission is coming from a single component, a flat-topped profile (from a shell like emission region) that is skewed either because of (for example) electron scattering or dust within the ejecta.

Although we can constrain the density in the region emitting [\ion{S}{ii}], {it is non-trivial to actually constrain the mass of ejected sulphur}, or of other heavy elements, since the emissivity is also very sensitive to the temperature.  
While it is difficult to firmly conclude a high mass of sulphur in iPTF14hls, we {propose} that the most likely origin of the strong [\ion{S}{ii}] emission is from nucleosynthesised material ejected in the core-collapse SN explosion.
{This proposal is based on the high density, high luminosity and intermediate velocity width of the line}.
 These lines were predicted to be 
from the core of massive-star ejecta, originating mainly from the Si and S-rich zone \citep{fransson1989}. 
 Although these models were powered by radioactive decay, a similar result is expected for powering by X-rays or UV radiation from circumstellar interaction or a magnetar, yet the relative line fluxes may change. Independent of the powering source, the high luminosity of these lines 
 {could indicate} processed material from the core.
This would then exclude a simple PPISN shell-shell scenario for which the core collapse is still to come. 

\subsection{Circumstellar Interaction from a core-collapse supernova}
\label{sec:csm}

Apart from CSM interactions related to PPI-eruptions as discussed above, the more general
CSM scenario has been invoked for many SNe~IIn. Although iPTF14hls showed early-time spectra similar to those of SNe~IIP, in terms of the LC there are many similarities to SNe~IIn. In particular, the slowly declining bolometric LC during the first $\sim 900$~d, as well as the steep decline (Fig.~\ref{bolo}), somewhat resemble those of SN 2010jl \citep[e.g.][]{Fransson2014}. 
For SN 2010jl, the early bolometric LC had a luminosity slope with a PL index 
$n=-0.54$ (20--320~d)
followed by a steeper decline of index $-3.4$ (320--1000~d). The first epoch is in accordance with simple CSM theory, whereas
the latter slope could have been due to a decrease in the CSM density.
For iPTF14hls the best-fit PL to the phase 600--900~d is $-3.6$, and after 950~d it steepens to $-13.5$.

While the late-time LC slope of iPTF14hls is thus much steeper than seen in SN 2010jl, 
it may not exclude the CSM scenario as we suggest it does for the accretion and magnetar models discussed above (unless one can abruptly shut off the central engine). 
CSM interaction can likely produce a large range of LC properties because of density variations related to the mass-loss history \cite[e.g.][]{nyholm} or due to geometry
\citep[see e.g.][their Fig.~18]{vanmarle}. The main arguments for CSM interaction in SN~2010jl were perhaps in the X-ray detections and in the spectral evolution, where this supernova differs from iPTF14hls. 

\subsubsection{X-rays}
\label{sec:xrays}

At the luminosity distance of iPTF14hls, the upper limit reported in Sect.~\ref{sec:swift} 
corresponds to a luminosity limit of $L_{\rm X} < 5.3 \times 10^{40}$
~erg~s$^{-1}$, at an epoch of $\sim 1194$ rest-frame days since discovery.
A long-lived CSM interacting SN like SN~2010jl had a detected X-ray luminosity of
$\sim 5 \times 10^{40}$~erg~s$^{-1}$ at this phase \citep{poonam}, 
so iPTF14hls was not more X-ray luminous than that, and at least fainter than the most X-ray bright SNe~IIn \citep[see][his Fig. 1]{dwarkadas14}. The X-ray to bolometric (optical) luminosity of iPTF14hls was thus $\lesssim 1.5$, while at the same epoch
SN~2010jl had $L_{\rm X}/L_{\rm opt} \approx 1$ \citep{poonam,jencson}. This may not exclude a CSM scenario, as pointed out by \cite{AS17}, but also does not support it. 
We note, however, that the X-ray limit for iPTF14hls discussed above was for a PL spectrum, as also discussed in A17. If we instead assume a thermal spectrum with a temperature of 20 keV, as observed for SN 2010jl \citep{poonam,ofek}, we get a limit of 
$1.0 \times 10^{-13}$~erg~cm$^{-2}$~s$^{-1}$ which corresponds to
a five-times less stringent upper limit in luminosity
($L_{\rm X} < 29 \times 10^{40}$~erg~s$^{-1}$).
{We are aware that Chandra X-ray observations as well as radio-observations have been obtained. These  may be able to further constrain the non-thermal emission from iPTF14hls.}

\subsubsection{Spectral evolution and narrow lines}
\label{sec:narrow}

Regarding the spectra, SN 2010jl showed a typical SN IIn spectral evolution dominated by narrow and intermediate-width Balmer lines, while iPTF14hls shows no signs of such emission lines for hundreds of days. 
\cite{AS17} argued for CSM interaction in iPTF14hls based on the emergence of a narrow line in 
their nebular spectrum. They furthermore noted the clear difference in line profiles of H$\alpha$ and O~I between day 600 (A17) and their spectrum at 1153~d. We discussed the spectral evolution of iPTF14hls in Sect.~\ref{sec:spectra}, and here we note in particular that the spectacular changes in the line profiles are well correlated in time with the dramatic evolution in the late-time LC. 
The appearance of metal lines from the inner core region is expected as the ejecta become transparent to electron scattering. The decreasing electron-scattering depth may be connected to the expansion of the ejecta as well as a decreasing state of ionisation. 

The very late-time spectra show asymmetric line profiles, perhaps not so different from what was seen in the Type IIn SN 1998S \cite[][see their Figs. 5 and 8]{pozzo}.
In SN 1998S, this asymmetry was interpreted as being caused by dust formation, whereas \citet[][their Fig. 10]{smith2015} discuss an asymmetric CSM disc.

The origin of the observed narrow H$\alpha$ in iPTF14hls is somewhat unclear. Such narrow lines may be expected in the CSM scenario where unshocked, circumstellar material is ionised by the shock radiation and recombines. The absence of such lines (and X-ray emission) was a key argument to not invoke a CSM mechanism in A17.

With the emergence of box-shaped line profiles and a narrow H$\alpha$ component, \cite{AS17} regarded the scenario with a previously hidden CSM scenario plausible. This conclusion was echoed by
\cite{mm18}, who presented a late-time (1210~d) spectrum of iPTF14hls and compared it with the Type IIn 
SN~2009ip, stating that the narrow lines are associated with nearby photoionised material.

\begin{figure}
\includegraphics[width=9cm,angle=-90]{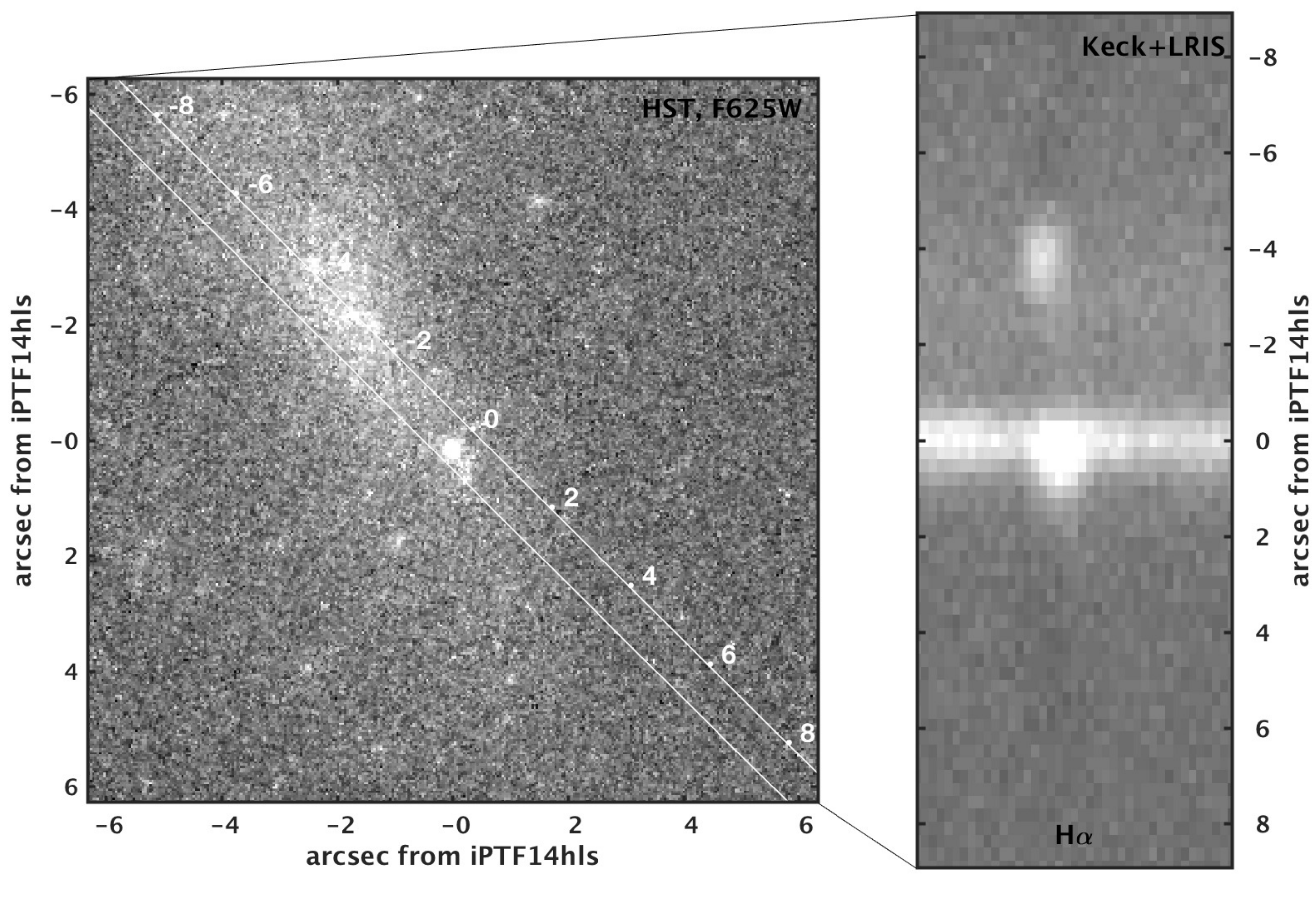}
\caption{\label{HII} Slit position marked (white lines) on the {\it HST} F625W image of iPTF14hls and the corresponding H$\alpha$ region of the two-dimensional spectrum obtained with 
Keck I on 4 January 2018. In this image we see two of the three \ion{H}{ii} regions along the slit 
positioned along the host galaxy.
One additional \ion{H}{ii} region is at the other end of the galaxy, while the ones here are close to the middle of the host, and one component of the narrow emission is located just at the SN position.  The position angle of the slit was $226^\circ$ in this last LRIS spectrum. In our earlier LRIS spectrum, no unresolved lines were detected. The slightly different position angle of the latter slit, $261^\circ$, explains why we do not pick up emission from the host galaxy, since the slit missed most of the galaxy at this angle. However, there is also no sign of narrow emission at the SN site at this phase, which is more puzzling. 
One arcsecond corresponds to 720 pc. Next to iPTF14hls an \ion{H}{ii} region can be discerned.}
\end{figure}

We have scrutinised the two-dimensional images of our Keck spectra to investigate the emergence of the narrow lines. 
In our final
spectrum (Keck/LRIS) these are very conspicuous, with unresolved emission from 
[\ion{O}{ii}]~$\lambda\lambda$3726, 3729, [\ion{O}{iii}]~$\lambda\lambda$4959, 5007, H$\beta$, H$\alpha$, and [\ion{S}{ii}] $\lambda\lambda$6716, 6731. These are typical H~II region emission lines, and we see emission from three locations along the slit positioned along the host galaxy. A representation of the slit around H$\alpha$ is shown in Fig.~\ref{HII}. 
The {\it HST} image also revealed several bright \ion{H}{ii} regions in the host galaxy, including a prominent \ion{H}{ii} region seen only 
$0\farcs5$ southwest of the SN. 
This may be the origin of the narrow emission lines seen in some of the spectra; see also \cite{AS17}.

In our earlier Keck/LRIS spectrum (+712 d), no unresolved lines were detected. The slightly different position angle of the latter slit
explains why we do not pick up emission from the host galaxy.
However, there is also no sign of narrow emission at the SN site at this phase, which is more puzzling. 
There is no narrow emission at this site in the Keck/HIRES spectrum from day 738, but it is present at 1081 days.
 We emphasise that even if the lines are narrow (unresolved), 
 the apparent lines and line strengths are typical of an \ion{H}{ii} region. For powering scenarios where the input is from shock emission, like in the current case for SN 1987A, one would perhaps expect a richer, and also higher ionisation (narrow) emission-line spectrum \cite[e.g.][]{groningsson}, and the same is true for a late-time input from a magnetar model. Both of these models would also have been further supported by strong X-ray emission, which we do not see.

\section{Summary}
\label{sec:summary}

The supernova iPTF14hls was already a unique object in the study of A17, and it has continued to evolve and deliver surprises in the nebular phase. 
From being almost impossible to explain (A17), a flood of papers followed, offering a multitude of different interpretations.

For example, the comprehensive model review by \cite{Woosley18} runs through a large number of models for CSM interaction, PPISNe, and magnetars. Without detailed comparisons to actual observations, 
\cite{Woosley18}
concludes that many models are versatile enough to produce a relatively luminous and long-lived LC for 600~d, although none of the many models accounts for all observed properties of iPTF14hls. Few of the numerous proposed models actually predicted the later observations presented here. In particular, the steep decline is difficult to explain either in the accretion model or in the magnetar model without some {ad hoc} mechanism, while similar drops have been observed for objects undergoing circumstellar interaction. 

This paper has thus provided a new suite of observations against which to test these models. 
{In particular, the very late (+1000 d) steep decline of the optical light curve is difficult to reconcile with the proposed central engine models. The lack of very strong X-ray emission, and the emergence of intermediate-width emission lines including [S II] that {could} originate from dense, processed material in the core of the supernova ejecta, are also key observational tests for existing and future models.} We hope our results will trigger a new effort towards modelling iPTF14hls.

\begin{acknowledgements}
We gratefully acknowledge support from the Knut and Alice
Wallenberg Foundation. Based on observations obtained with the Samuel Oschin 48-inch telescope and the 60-inch telescope at the Palomar Observatory as part of the intermediate Palomar Transient Factory (iPTF) project, a
scientific collaboration among the California Institute of Technology,
Los Alamos National Laboratory, the University of Wisconsin
(Milwaukee), the Oskar Klein Center, the Weizmann Institute of Science,
the TANGO Program of the University System of Taiwan, and the Kavli
Institute for the Physics and Mathematics of the Universe. 
The Oskar Klein Centre is funded by the Swedish Research Council. 
This work is partly based on observations made with DOLoRes@TNG, and 
on observations made with the Nordic Optical Telescope, operated by NOTSA at IAC using
ALFOSC, which is provided by the IAA. 
Some of the data presented herein were obtained at the W.M. Keck
Observatory, which is operated as a scientific partnership among the
California Institute of Technology, the University of California, and
NASA; the observatory was made possible by the generous financial
support of the W.M. Keck Foundation. Some of the
Keck observations were partially supported by Northwestern University and the Center for Interdisciplinary Exploration and Research in Astrophysics (CIERA).
Support for I. Arcavi was provided by NASA through the Einstein Fellowship Program, grant PF6-170148. R.L. acknowledge GROWTH. Research funding to I. Andreoni is provided by the Australian Astronomical Observatory. A.V.F. has been supported by the Christopher R. Redlich Fund, the TABASGO Foundation, and the Miller Institute for Basic Research in Science (U.C. Berkeley) where he is a Miller Senior Fellow.
His work was conducted in part at the Aspen Center for Physics, which is 
supported by U.S. National Science Foundation grant PHY-1607611; he thanks the Center for its hospitality during the supermassive black holes workshop in June and July 2018.  X. W. is supported by the National Natural Science Foundation of China
 (NSFC grants  11325313 and 11633002). Special thanks to A. Bostroem and W. Zheng for help with Keck observations

\end{acknowledgements}

\bibliographystyle{aa}

\onecolumn

\begin{deluxetable}{|cc|cc|cc|cc|cc|}
\tabletypesize{\scriptsize}
\tablewidth{0pt}
\tablecaption{Late-time optical photometry of iPTF14hls.\label{tab:phot}}
\tablehead{
\colhead{JD}&
\colhead{$B$}&
\colhead{JD}&
\colhead{$g$}&
\colhead{JD}&
\colhead{$V$}&
\colhead{JD}&
\colhead{$r$}&
\colhead{JD}&
\colhead{$i$}\\
\colhead{(days)}&
\colhead{(mag)}&
\colhead{(days)}&
\colhead{(mag)}&
\colhead{(days)}&
\colhead{(mag)}&
\colhead{(days)}&
\colhead{(mag)}&
\colhead{(days)}&
\colhead{(mag)}}
\startdata
\multicolumn{10}{c}{P60 SEDM}\\
&  &   2457660.995 & 19.930(0.030) & & &  2457671.905 & 19.320(0.070)  &  2457660.993 &  19.260(0.030)\\
&  &   2457671.908 & 20.320(0.090) & & &  2457687.874 & 19.540(0.060)  &  2457671.907 &  19.530(0.090)\\
&  &   2457687.877 & 20.140(0.050) & & &  2457700.916 & 19.540(0.040)  &  2457687.875 &  19.540(0.080)\\
&  &   2457700.920 & 20.180(0.040) & & &  2457706.813 & 19.580(0.160)  &  2457700.918 &  19.520(0.050)\\
&  &   2457715.882 & 20.170(0.050) & & &  2457715.879 & 19.680(0.040)  &  2457706.814 &  19.500(0.160)\\
&  &   2457730.900 & 20.160(0.040) & & &  2457730.897 & 19.600(0.040)  &  2457730.898 &  19.730(0.060)\\
&  &   2457736.791 & 19.860(0.220) & & &  2457736.788 & 19.610(0.200)  &  2457736.790 &  19.720(0.170)\\
&  &   2457742.730 & 20.340(0.120) & & &  2457742.726 & 19.690(0.110)  &  2457742.728 &  19.790(0.120)\\
&  &   2457750.737 & 20.070(0.200) & & &  2457757.905 & 19.910(0.070)  &  2457770.769 &  19.960(0.190)\\
&  &   2457757.909 & 20.300(0.100) & & &  2457770.767 & 19.890(0.110)  &  2457789.719 &  19.790(0.060)\\
&  &   2457770.771 & 20.040(0.130) & & &  2457789.718 & 19.750(0.070)  &  2457797.958 &  20.000(0.120)\\
&  &   2457789.721 & 20.220(0.060) & & &  2457797.956 & 19.850(0.120)  &                              \\
\hline

\multicolumn{10}{c}{LCO 1~m}\\
 2457660.965 & 19.924(0.044) & 2457660.979 & 19.680(0.024) & 2457660.972 & 19.529(0.044) &   2457660.986 & 19.066(0.025) &        2457660.992 & 19.405(0.090) \\ 
 2457665.952 & 19.914(0.049) & 2457665.966 & 19.774(0.024) & 2457665.960 & 19.698(0.049) &   2457665.974 & 19.085(0.030) &        2457665.980 & 19.402(0.063) \\ 
 2457672.961 & 19.780(0.056) & 2457672.974 & 19.595(0.029) & 2457672.968 & 19.486(0.059) &   2457672.982 & 18.985(0.033) &        2457680.949 & 19.682(0.138) \\ 
 2457678.961 & 19.714(0.097) & 2457680.929 & 19.876(0.066) & 2457680.921 & 19.757(0.090) &   2457680.939 & 19.335(0.047) &        2457678.998 & 19.645(0.089) \\ 
 2457692.968 & 20.141(0.037) & 2457678.980 & 19.927(0.059) & 2457678.971 & 19.850(0.099) &   2457678.989 & 19.219(0.041) &        2457676.996 & 19.494(0.073) \\ 
 2457687.919 & 20.100(0.039) & 2457676.978 & 19.745(0.070) & 2457676.970 & 19.629(0.085) &   2457676.988 & 19.262(0.041) &        2457693.004 & 19.731(0.047) \\ 
 2457711.950 & 20.077(0.112) & 2457668.957 & 19.839(0.059) & 2457668.950 & 19.630(0.096) &   2457668.964 & 19.132(0.077) &        2457687.956 & 19.719(0.065) \\ 
             &               & 2457692.987 & 19.938(0.025) & 2457692.978 & 19.809(0.037) &   2457706.874 & 19.312(0.120) &                    &               \\
             &               & 2457687.938 & 19.908(0.021) & 2457687.929 & 19.869(0.049) &   2457692.996 & 19.360(0.024) &                    &               \\
             &               & 2457711.966 & 19.871(0.068) &             &               &   2457687.948 & 19.403(0.022) &                    &               \\
             &               &             &               &             &               &   2457711.976 & 19.471(0.077) &                    &               \\
\hline  

\multicolumn{10}{c}{LCO 2~m}\\
 2457736.017 & 19.520(0.364) &  2457743.058 & 20.080(0.087) &  2457743.050 & 19.918(0.086) &      2457743.068 & 19.522(0.021) &      2457743.075 & 19.688(0.041) \\
 2457743.039 & 20.023(0.195) &  2457748.023 & 19.931(0.043) &  2457748.015 & 19.919(0.051) &      2457748.032 & 19.565(0.024) &      2457748.040 & 19.590(0.039) \\
 2457748.016 & 20.185(0.097) &  2457753.910 & 20.188(0.094) &  2457753.901 & 19.763(0.101) &      2457753.919 & 19.466(0.049) &      2457753.924 & 19.564(0.125) \\
 2457755.973 & 19.922(0.066) &  2457755.991 & 19.711(0.042) &  2457761.931 & 19.982(0.127) &      2457756.000 & 19.621(0.017) &      2457756.006 & 19.978(0.051) \\
 2457771.840 & 20.046(0.193) &  2457764.921 & 19.492(0.447) &  2457771.864 & 20.116(0.115) &      2457771.925 & 19.679(0.048) &      2457761.942 & 19.946(0.115) \\
 2457772.909 & 20.145(0.095) &  2457771.916 & 19.914(0.079) &  2457772.919 & 19.951(0.056) &      2457772.936 & 19.654(0.022) &      2457772.942 & 19.958(0.036) \\
 2457778.010 & 20.473(0.166) &  2457772.927 & 20.031(0.045) &  2457778.019 & 20.085(0.069) &      2457778.036 & 19.646(0.029) &      2457778.044 & 19.904(0.050) \\
 2457784.081 & 19.694(0.085) &  2457778.027 & 20.003(0.039) &  2457784.090 & 20.053(0.098) &      2457784.107 & 19.639(0.040) &      2457784.115 & 19.796(0.074) \\
 2457789.861 & 19.758(0.134) &  2457784.098 & 19.895(0.068) &  2457789.873 & 19.979(0.088) &      2457789.890 & 19.319(0.080) &      2457811.811 & 19.877(0.067) \\
 2457795.075 & 20.573(0.388) &  2457789.881 & 20.220(0.062) &  2457811.788 & 19.842(0.122) &      2457799.052 & 19.754(0.076) &      2457816.765 & 20.071(0.080) \\
 2457827.843 & 20.163(0.195) &  2457795.092 & 19.794(0.283) &  2457816.741 & 19.865(0.098) &      2457811.804 & 19.638(0.040) &      2457827.877 & 20.121(0.089) \\
 2457839.877 & 20.124(0.102) &  2457799.041 & 20.026(0.090) &  2457827.853 & 19.961(0.078) &      2457816.758 & 19.713(0.042) &      2457839.961 & 20.074(0.096) \\
 2457850.778 & 20.280(0.222) &  2457811.794 & 19.883(0.049) &  2457839.936 & 19.643(0.102) &      2457827.870 & 19.765(0.028) &      2457904.795 & 20.125(0.155) \\
 2457889.806 & 20.399(0.201) &  2457816.749 & 19.953(0.062) &  2457850.787 & 20.364(0.136) &      2457839.953 & 19.662(0.067) &                  &               \\
                 &               &  2457827.861 & 20.188(0.115) &  2457889.815 & 19.854(0.091) &      2457889.833 & 19.750(0.055) &                  &               \\
                 &               &  2457839.944 & 20.463(0.170) &  2457904.770 & 19.548(0.249) &      2457904.788 & 19.785(0.075) &                  &               \\
                 &               &  2457889.823 & 20.078(0.096) &              &               &                  &               &                  &               \\
\hline

\multicolumn{10}{c}{NOT}\\
        &    &      2458044.713 & 21.466(0.058) &  & &       2458044.721 & 21.715(0.091) &         2458044.729 & 21.658(0.055) \\
        &    &      2458056.665 & 21.601(0.069) &  & &       2458056.673 & 21.803(0.094) &         2458056.681 & 21.783(0.088) \\
        &    &      2458067.696 & 21.708(0.191) &  & &       2458067.705 & 21.759(0.180) &         2458075.650 & 21.959(0.110) \\
        &    &      2458075.633 & 21.777(0.085) &  & &       2458075.641 & 22.026(0.109) &         2458093.692 & 22.195(0.248) \\
        &    &      2458093.675 & 21.961(0.241) &  & &       2458093.684 & 22.177(0.214) &                     &               \\
        &    &      2458128.576 & 22.530(0.098) &  & &                   &               &                     &               \\
        &    &      2458146.609 & 22.607(0.266) &  & &                   &               &                     &               \\
\hline  

\multicolumn{10}{c}{TNG}\\
 & &     2458077.598 & 21.664(0.076) & & & 2458077.578 & 21.823(0.117) & 2458077.588 & 21.836(0.099) \\
 & &     2458129.637 & 22.403(0.121) & & & 2458129.659 & 22.678(0.172) & 2458129.648 & 22.718(0.213) \\
 & &     2458171.393 & 23.187(0.234) & & & 2458171.379 & 23.055(0.168) & 2458171.406 & 22.850(0.296) \\
 & &     2458201.485 & 23.308(0.259) & & & 2458201.510 & 23.313(0.163) & 2458201.493 & 23.137(0.254) \\
\hline 
\multicolumn{10}{c}{HST*}\\
&  &   2458108.325 & {22.383(0.015)} & & &  2458108.343 & { 22.470(0.018)}  &   &  \\ 
\enddata
\tablenotetext{*}{F475W and F625W approximately correspond to $g$ and $r$ filters.}
\end{deluxetable}

\begin{deluxetable}{lccccc}
\tablewidth{0pt}
\tablecaption{\textit{Swift} observations of iPTF14hls \label{tab:swift}}
\tablehead{
\colhead{Start Date}&
\colhead{Phase\tablenotemark{a}}&
\colhead{XRT Exposure time}&
\colhead{Count rate limit}&
\colhead{Flux limit}&
\colhead{Luminosity limit}\\
\colhead{(UT)}&
\colhead{(days)}&
\colhead{(s)}&
\colhead{($10^{-3}$ counts~s$^{-1}$)}&
\colhead{(10$^{-14}$~erg~cm$^{-2}$~s$^{-1}$)}&
\colhead{(10$^{41}$~erg~s$^{-1}$)}}
\startdata
           2015-05-23 01:10:59  & +234.45 & 4941.9      & 0.87 & 3.1 & 0.91 \\ 
           2017-06-14 02:37:57  & +962.47 & 4488.1      & 0.92 & 3.4 & 0.99 \\
           2017-11-02 23:58:57  & +1099.6 & 4781.4      & 0.94 & 3.4 & 0.99 \\
           2017-11-07 01:31:57  & +1103.6 & 3079.0      & 2.92 & 10.6 & 3.1 \\
           2017-12-19 03:39:57  & +1144.3 & 4280.0      & 1.45 & 5.3 & 1.5 \\
           2018-02-07 07:17:56  & +1192.7 & 6075.2      & 0.75 & 2.7 & 0.79 \\
           2018-02-08 08:42:57  & +1193.8 & 1995.8      & 1.96 & 7.1 & 2.1 \\
           2018-02-11 00:30:57  & +1196.3 & 3783.5      & 1.22 & 4.4 & 1.3 \\
\enddata
\tablenotetext{*}{The 90\% upper limits are given for the 
0.3--10.0
keV range. The flux limits further assume a power-law spectrum with photon index of $\Gamma
= 2$ and a Galactic hydrogen column density of $1.4\times10^{20}$
cm$^{-2}$.}
\tablenotetext{a}{Rest-frame days from discovery.}
\end{deluxetable}

\begin{deluxetable}{ccccccc}
\tabletypesize{\scriptsize}
\tablewidth{0pt}
\tablecaption{Late-time optical spectroscopy of iPTF14hls \label{tab:spectra}}
\tablehead{
\colhead{Date (UT)}&
\colhead{JD-2,457,000}&
\colhead{Phase\tablenotemark{a}}&
\colhead{Telescope}&
\colhead{Instrument}&
\colhead{FWHM}&
\colhead{Range}\\
\colhead{}&
\colhead{(days)}&
\colhead{(days)}&
\colhead{}&
\colhead{}&
\colhead{(\AA)}&
\colhead{(\AA)}}
\startdata
2016-09-28   &        660.13 &   +712.59    &  Keck~I  &  LRIS     & 6   & 3162--10274   \\
2016-09-30   &        661.73 &   +714.13    &  NOT     &  ALFOSC   & 17  & 3629--9713    \\
2016-10-08   &        670.09 &   +722.22    &  FTN     &  FLOYDS   & 16  & 3301--9301    \\
2016-10-14   &        676.08 &   +728.01    &  FTN     &  FLOYDS   & 16  & 4800--9300    \\
2016-10-25   &        687.10 &   +738.66    &  Keck~II &  DEIMOS   & 3   & 4440--9638    \\
2016-10-25   &        687.15 &   +738.71    &  Keck~II &  DEIMOS   & 3   & 4771--7428    \\
2016-11-08   &        701.11 &   +752.20    &  FTN     &  FLOYDS   & 16  & 3999--9300    \\
2016-11-24*  &        717.12 &   +767.68    &  FTN     &  FLOYDS   & 16  & 4800--10000   \\
2016-11-29   &        721.71 &   +772.12    &  NOT     &  ALFOSC   & 17  & 3833--9710    \\
2016-12-13*  &        736.06 &   +785.99    &  FTN     &  FLOYDS   & 16  & 5000--9300    \\
2017-02-28   &        813.47 &   +860.83    &  NOT     &  ALFOSC   & 17  & 3334--9715    \\
2017-03-26   &        838.52 &   +885.04    &  NOT     &  ALFOSC   & 17  & 3537--9710    \\
2017-04-04   &        848.41 &   +894.61    &  TNG     &  DOLORES  & 14  & 3282--10462   \\
2017-04-24   &        867.50 &   +913.06    &  TNG     &  DOLORES  & 14  & 3401--8566    \\
2017-05-26   &        900.42 &   +944.89    &  NOT     &  ALFOSC   & 17  & 3464--9708    \\
2017-05-30   &        903.78 &   +948.14    &  Keck~I  &  LRIS     & 6   & 3150--10237   \\
2017-10-14   &       1041.12 &   +1080.91   &  Keck~II &  DEIMOS   & 3   & 4450--9634    \\
2017-10-26   &       1052.70 &   +1092.10   &  NOT     &  ALFOSC   & 17  & 3817--9636    \\
2017-11-13   &       1071.04 &   +1109.83   &  Keck~II &  DEIMOS   & 3   & 4320--9904    \\
2017-11-18   &       1076.05 &   +1114.67   &  Keck~I  &  LRIS     & 6   & 3138--10233   \\
2017-12-02   &       1089.62 &   +1127.79   &  NOT     &  ALFOSC   & 17  & 3432--9714    \\
2017-12-29   &       1116.57 &   +1153.85   &  NOT     &  ALFOSC   & 17  & 3780--9720    \\   
2018-01-14   &       1132.93 &   +1169.66   &  Keck~I  &  LRIS     & 6   & 3069--10266   \\
\enddata
\tablenotetext{a}{Rest-frame days from discovery.}
\tablenotetext{*}{Not shown in Fig.~\ref{specseq}.}
\end{deluxetable}

\begin{deluxetable}{lcc|cc}
\tablewidth{0pt}
\tablecaption{Two Gaussian components of H$\alpha$, $[\ion{O}{i}]~\lambda\lambda$6300, 6364, and  $[\ion{S}{ii}]~\lambda\lambda$4069, 4076.\label{tab:linecomp}}
\tablehead{
\colhead{Line}&
\colhead{$\Delta v$ (red)}&
\colhead{FWHM (red) }&
\colhead{$\Delta v$ (blue)}&
\colhead{ FWHM (blue) }\\
\colhead{}&
\colhead{(km~s$^{-1}$)}&
\colhead{(km~s$^{-1}$)}&
\colhead{(km~s$^{-1}$)}&
\colhead{(km~s$^{-1}$)}}
\startdata
\multicolumn{5}{c}{\bf +1081~d}\\
         H$\alpha$       & 734(19) & 1630(32) &  -885(14)  & 1564(21)  \\
         $[\ion{O}{i}]$  & 636(67) &   2226(120) &  -1242(29) & 1400(52)  \\
         \hline
         \multicolumn{5}{c}{\bf +1110~d}\\
         H$\alpha$       & 946(64)  & 1398(115) &   -810(57) & 1927(124)  \\
         $[\ion{O}{i}]$  & 156(134) &  3534(659) &  -1774(51) &   1300(101)  \\
     \hline
         \multicolumn{5}{c}{\bf +1115~d}\\
         H$\alpha$       & 771(50)  & 1441(90) &   -931(28) & 1566(50)  \\
         $[\ion{O}{i}]$  & 444(182) &  2177(299) &  -1260(58) &   1242(103)  \\
         \hline
         \multicolumn{5}{c}{\bf +1170~d}\\
         H$\alpha$       & 701(88)  & 1655(163) &  -997(43)  & 1395(71)  \\
         $[\ion{O}{i}]$  & 565(154) & 1751(350) &  -1774(40) &   1286(87)  \\
         $[\ion{S}{ii}]$ & 170(175) & 2508(250) &  -1550(38) & 1074(99)  \\

\enddata
\end{deluxetable}

\end{document}